\documentclass[a4paper,11pt]{article}
\pdfoutput=1 
\usepackage{jheppub} 
\usepackage{enumerate}
\usepackage[T1]{fontenc} % if needed
\usepackage[utf8]{inputenc} 
\usepackage[bottom]{footmisc}

\makeatletter
\newcommand\footnoteref[1]{\protected@xdef\@thefnmark{\ref{#1}}\@footnotemark}
\makeatother

\renewcommand{\L}{\mathcal{L}}

\renewcommand{\l}{\lambda}
\renewcommand{\d}{\partial}

\renewcommand{\a}{\alpha}
\renewcommand{\b}{\beta}

\title{\boldmath Gravity, Scale Invariance and the Hierarchy Problem}

\author[a]{Mikhail Shaposhnikov,}
\author[a,b]{Andrey Shkerin}

\affiliation[a]{Institute of Physics, Ecole Polytechnique F\'ed\'erale de Lausanne (EPFL),\\1015 Lausanne, Switzerland}
\affiliation[b]{Institute for Nuclear Research of the Russian Academy
of Sciences,\\60th October Anniversary prospect 7a, 117312, Moscow,
Russia}

\emailAdd{mikhail.shaposhnikov@epfl.ch}
\emailAdd{andrey.shkerin@epfl.ch}

\abstract{Combining the quantum scale invariance with the absence of new degrees of freedom above the electroweak scale leads to stability of the latter against perturbative quantum corrections. Nevertheless, the hierarchy between the weak and the Planck scales remains unexplained. We argue that this hierarchy can be generated by a non-perturbative effect relating the low energy and the Planck-scale physics. The effect is manifested in the existence of an instanton configuration contributing to the vacuum expectation value of the Higgs field. We analyze such configurations in several toy models and in a phenomenologically viable theory encompassing the Standard Model and General Relativity in a scale-invariant way. Dynamical gravity and a non-minimal coupling of it to the Higgs field play a crucial role in the mechanism. }

\begin{document} 
\maketitle
\flushbottom

\section{Introduction}
\label{Sec:intro}

As suggested by the long quest for unification of fundamental interactions, it is natural to search for principles relating phenomena that occur at very different energy scales. To an underlying theory unifying diverse physical processes we assign a task to explain possible large differences in the measured fundamental quantities. One of the most striking differences, which has been a source of new ideas in particle physics for decades, is manifested in the ratio of the Fermi constant $G_F$, that sets the weak interaction scale, to the Newton constant $G_N$ determining the gravitational force strength,\footnote{For illustrative purposes, here we write explicitly the Planck constant $\hbar$ and the speed of light $c$. Everywhere further we work in natural units $\hbar=c=1$.}
\begin{equation} \label{TheProblem}
\dfrac{G_F\hbar^2}{G_N c^2}\sim 10^{33} \,.
\end{equation}
It is tempting to speculate that some deep reason for the big number to appear in this relation may be hidden in a yet unknown theory encompassing the Standard Model (SM) and General Relativity.

At the classical level, the ratio (\ref{TheProblem}) represents one face of the problem. Another aspect of it appears when we adopt the quantum field theory framework. It originates from properties of the Higgs field through the vacuum expectation value (vev) of which the Fermi constant is defined. As it was realized long ago in studies of Grand Unified Theories, whenever new physics comes about with heavy degrees of freedom (dof) activating at some mass scale $M_X$, the heavy particle's loops are expected to produce an additive correction to the Higgs mass $m_H$ \cite{Wilson:1970ag,Gildener:1976ai,Susskind:1978ms,Weinberg:1978ym,Vissani:1997ys},\footnote{We assume that the new particles are coupled sufficiently strongly to the Higgs field.}
\begin{equation} \label{HiggsMass_GUT}
\delta m_{H,\,X}^2\sim M_X^2\,.
\end{equation}
As soon as $M_X$, if exists, is much larger than the observed value of $m_H$, eq. (\ref{HiggsMass_GUT}) implies either a fine-tuning between various contributions to the Higgs mass or a mechanism of systematic suppression of those contributions. This puzzling fact about the SM Higgs field is known as the Electroweak (EW) hierarchy problem. If one now treats the EW and gravitational forces within the quantum field theory framework, then one must include quantum gravity loop corrections to the Higgs mass. The naive power counting argument suggests these corrections to be of the order of the Planck mass,
\begin{equation} \label{HiggsMass_Grav}
\delta m_{H,\,grav.}^2\sim M_P^2 \,.
\end{equation}
The validity of this estimation can be doubted by the observation that, unlike $M_X$, the Planck mass defines an interaction scale rather than a new particle's mass scale (see, e.g., \cite{Shaposhnikov:2007nj}).\footnote{In fact, this observation can well be applied to the case of new physics much below the Planck scale. For example, in \cite{Karananas:2017mxm} an interpretation of the gauge coupling unification scale was proposed, which is not related to any new particle threshold; see also \cite{Giudice:2016yja}.} Moreover, at the energies close to $M_P$ gravity enters the strong-coupling regime where estimations based on perturbation theory loose the predictive power. Nevertheless, if we admit eq. (\ref{HiggsMass_Grav}), then the observed difference in the interaction strengths (\ref{TheProblem}) either requires a remarkable balance between the EW and the Planck scale physics, or it is an indication of specific properties of quantum gravity at strong coupling that result in the absence of the quadratic corrections to $m_H$, see \cite{Giudice:2008bi,Giudice:2013nak} for reviews of the problem.

The hierarchy problem was addressed in literature many times and from various perspectives. The list of proposals dealing with the problem by introducing a new physics close to the EW scale includes supersymmetry, composite Higgs theories (for reviews see \cite{Feng:2013pwa,Bellazzini:2014yua} correspondingly), extra dimensions \cite{ArkaniHamed:1998rs,Randall:1999ee}. The parameter spaces of the models extending the SM at the TeV scale are subject to constraints provided, in particular, by the LHC data. These constraints force such theories to be fine-tuned in order to remain compatible with experiment \cite{Feng:2013pwa,Wells:2013tta,Giudice:2013nak,Barnard:2015ryq}. More recent proposals attempt to overcome this issue \cite{Farina:2013mla,Arvanitaki:2016xds,Pelaggi:2017wzr,Salvio:2017qkx}. Some of them suggest mechanisms of generation of exponentially small couplings to the Higgs field \cite{Giudice:2016yja}, or rely on a specific dynamics of the latter during the cosmological evolution \cite{Graham:2015cka}. 

Regardless the particular content of a model extending the SM at high energies, a common approach to the hierarchy problem lies within the effective field theory framework. The latter implies that the low energy description of Nature, provided by the SM, can be affected by an unknown UV physics only through a finite set of parameters. Two of them -- the mass of the Higgs boson and the cosmological constant -- are most sensitive to the scale and to the dynamics of physics beyond the SM, being quadratically and quartically divergent. In ``natural''  theories the quadratically divergent UV contributions to the Higgs mass are  eliminated by  introducing new physics right above the Fermi scale.  It is this naturalness principle that is seriously questioned now in light of the absence of signatures of new physics at the TeV scale \cite{Giudice:2013nak}.  While some parameter regions of the theories with $M_X\sim 1\,\text{TeV}$ still survive at the price of a moderate fine-tuning, a relatively radical step would be to suggest that the UV physics can affect the low energy behavior in a way that is not captured by the perturbation theory. Going back to the ratio (\ref{TheProblem}), this would imply the existence of a non-perturbative effect linking the scales separated by $17$ orders of magnitude.

The idea of some principle that can shape the behavior of a theory at very different energy scales is not novel to particle physics. For example, it is tempting to use such kind of reasoning when investigating a probable (near-)degeneracy of the minima of the Renormalization Group (RG) improved SM Higgs potential, which is supported by the recent measurements of the Higgs and top quark masses \cite{Bezrukov:2012sa,Buttazzo:2013uya}. A possible mechanism that makes the form of the potential special and, hence, predicts the values of the low energy parameters, can manifest itself in a number of ways. For example, in \cite{Froggatt:1995rt} bounds on the Higgs and top quark masses were put based on the principle of multiple point criticality \cite{Bennett:1993pj}, while in \cite{Shaposhnikov:2009pv} the prediction of $m_H$ was made, guiding by an asymptotic safety of gravity \cite{Weinberg:2009wa}. Inspired by these ideas, in this paper we make an attempt to resolve the problem (\ref{TheProblem}) by looking for an inherently non-perturbative effect relating the weak and the Planck scales.

Non-perturbative physics provides natural tools to establish links between the low energy and the high energy regimes of a theory. Perhaps, the most striking example of such a link, which strongly interferes phenomenology, is revealed in studying the EW vacuum decay. Indeed, it is known that, depending on the structure of UV operators added to the SM at large energy scales, the decay rate of the EW vacuum can be changed drastically compared to the pure SM case \cite{Branchina:2014rva}. Hence, having observed the sufficiently long-lived Universe, one can make certain predictions about the physics complementing the SM at high energies (for a review see \cite{Bezrukov:2014ina} and references therein).

The formulation of the hierarchy problem necessarily implies at least two scales in game. Considered isolated, the SM does not possess the problem due to the absence of thresholds with the energies above the EW scale, with which the Higgs mass is to be compared \cite{Bardeen:1995kv}.\footnote{Here we leave aside an issue with the Landau pole in the scalar self-coupling.} But as soon as gravity is embedded into the quantum field theory framework, one high energy scale appears inevitably, raising the question about the origin of the big number in the r.h.s. of eq. (\ref{TheProblem}).\footnote{ The fact that a solution of the hierarchy problem may require a theory of unification with gravity was pointed out in \cite{Wetterich:1983bi}, see also \cite{Wetterich:2011aa}.  } 
Nonetheless, the no-scale scenario looks attractive \cite{Fujii:1974bq,Fujii1975,Amit:1984ri,Rabinovici:1987tf,Wetterich:1987fm}, and motivates to search for the models which, alongside with incorporating the SM and gravity, do not contain dimensional parameters at the classical level \cite{Zee:1978wi,Smolin:1979uz,Buchmuller:1988cj,Shaposhnikov:2008xb,GarciaBellido:2011de,Bezrukov:2012hx,Rubio:2014wta,Ferreira:2016vsc,Karananas:2016kyt,Ferreira:2016wem,Ferreira:2016kxi}. The advantage of this approach is that scale invariance and the absence of new heavy particles can protect the Higgs mass from large radiative corrections, thus making its value natural according to the 't Hooft definition \cite{tHooft:1979rat}. This is a step forward in a solution of the hierarchy problem, although the big difference between the Fermi and the Planck scales remains unexplained. Here we do not discuss the possibility for the mass term to appear in the RG-improved Higgs potential via Coleman-Weinberg mechanism \cite{Coleman:1973jx,Weinberg:1978ym}. This scenario can indeed be realized, but in the SM it leads to the Higgs and the top quark masses being far from those observed experimentally; for discussion, see, e.g., \cite{Shaposhnikov:2018xkv} and references therein. To study non-perturbative phenomena that can possibly affect the Higgs mass, it would be useful to have a theory in which the corrections coming at the perturbative level are suppressed, and one can achieve this by the means of scale symmetry and by requiring that no heavy dof appear beyond the SM.

In the scale invariant (SI) framework, the Planck mass appears as a result of a spontaneous breaking of the scale symmetry. The aim of this paper is to argue that gravitational effects can generate non-perturbatively a new scale, associated with the classically zero vev of a scalar field. Dynamical gravity and global scale symmetry are important ingredients of a theory admitting this non-perturbative mechanism. The former ensures the existence of euclidean classical configurations of a special type --- the singular instantons --- that contribute to the vev of the scalar field. The latter can protect the vev from large radiative corrections, provided that the scalar sector of a theory is additionally invariant under constant shifts of the field responsible for generating the Planck scale \cite{Shaposhnikov:2008xi}, see section \ref{Sec:Pheno} for detail. Our goal is to find if it is possible, in a particular class of theories, to make the new scale much smaller than $M_P$, in which case the hierarchy of scales emerges.
 
The existence of the desired instanton configuration relies on a specific structure of a theory in the high energy and large field limits. In this paper, we investigate this structure by the means of simple SI models containing the gravitational and scalar dof, that mimic the Higgs-gravity sector of the theory we are eventually interested in. To apply the results of the non-perturbative analysis to the actual hierarchy problem (\ref{TheProblem}), it is necessary to have a theory which is compatible with the models on which the mechanism is tested and is consistent with observations and experiment. A good candidate for such theory is the Higgs-Dilaton model \cite{Fujii:1974bq,Wetterich:1987fm,GarciaBellido:2011de,Wetterich:2011aa,Bezrukov:2012hx}. We will show how eq. (\ref{TheProblem}) is reproduced in a certain modification of this model, which preserves all phenomenological consequences of the original theory.

Of course, the absence of an explicit UV completion of gravity engenders irremovable ambiguities in our analysis. The SI framework and the requirement of having a phenomenologically viable low energy limit reduce partially this ambiguity. The resulting amount of possibilities for choosing a particular model for the analysis is, however, still too large.\footnote{For example, the Higgs-gravity sector of a theory under investigation can be governed by the Horndeski Lagrangian or its extensions \cite{Horndeski:1974wa,Nicolis:2008in,Langlois:2015cwa}.} In this paper, we focus on some possible examples of models in which the suggested mechanism of the exponential suppression of the Planck scale due to instantons exists. We do not intend to perform an extensive survey of all possible examples. Nor do we intend to argue that the toy model chosen to illustrate the mechanism can indeed be consistently embedded into the UV complete theory of gravity. Note, however, that eq. (\ref{TheProblem}) can be viewed as an argument in favor of those properties of a UV theory, that support the existence of the suppression mechanism.

In this paper, we follow the ideas of  the work \cite{Shaposhnikov:2018xkv}, where the non-perturbative mechanism of generation of the scalar field vev was studied in scalar-tensor theories with an explicit breaking of global scale symmetry in the gravitational sector at low energies. The conditions for the successful implementation of the mechanism found there are similar to those discussed in the current work.

The paper is organized as follows. In section \ref{Sec:Idea} we describe a general idea of how to capture instanton contributions to the vev of a scalar field. In section \ref{Sec:D} we study the toy SI model containing one scalar dof coupled to gravity in a non-minimal way. The advantage of this model is that its euclidean classical configurations can be found in an analytic form. We describe the important properties of these configurations, which they share with the instantons arising in a more complicated setting.

In section \ref{Sec:DD} we introduce a class of SI models of gravity with two scalar dof. We specify the properties a model should obey in order to be compatible with a phenomenologically viable theory encompassing the SM and General Relativity. We then study in detail classical configurations arising in the chosen class of models. We identify the features which support the mechanism of generating the hierarchy of scales via the instantons. In section \ref{Sec:Pheno} the results of the analysis are applied to the EW hierarchy problem. There, we first outline the Higgs-Dilaton model and propose its modification in the limit of large magnitudes and momenta of the Higgs field. We then demonstrate that the instantons can contribute to the vev of the Higgs field so that to reproduce eq. (\ref{TheProblem}). In section \ref{Sec:Disc} we discuss our findings and conclude.

\section{Outline of the idea}
\label{Sec:Idea}

We begin with providing a general idea of the method that allows to capture non-perturbative gravitational contributions to a one-point correlation function of a scalar field. Consider the theory containing a real scalar field $\varphi$ of a unit mass dimension, the metric field $g_{\mu\nu}$ and, possibly, other dof which we denote collectively by $\mathcal{A}$. In the euclidean signature, the (time-independent, spatially homogeneous) vev of $\varphi$ is evaluated as\footnote{Here and below we work with euclidean formulation of theories, without indicating this explicitly. We will comment on this later in this section.}
\begin{equation}\label{Idea:vev}
\langle\varphi\rangle=Z^{-1}\int\mathcal{D}\varphi\mathcal{D}g_{\mu\nu}\mathcal{D}\mathcal{A}\:\varphi(0) e^{-S} \; ,
\end{equation}
where $Z$ denotes the partition function,
\begin{equation}\label{Idea:part}
Z=\int\mathcal{D}\varphi\mathcal{D}g_{\mu\nu}\mathcal{D}\mathcal{A}\: e^{-S} \; ,
\end{equation}
and $S$ is the euclidean action of the theory. If the theory admits the classical ground state of the form $\varphi=0$, $g_{\mu\nu}=\delta_{\mu\nu}$, the numerator in eq. (\ref{Idea:vev}) can be computed by the means of the standard perturbation theory. Let us instead attempt to reorganize it by exponentiating the scalar field variable in the region of large magnitudes of the latter,
\begin{equation}\label{Idea:NewVar}
\varphi\rightarrow\varphi_0 e^{\bar{\varphi}} \; , ~~~ \varphi\gtrsim\varphi_0 \; ,
\end{equation}
where by $\varphi_0$ we understand an appropriate scale of the theory. The corresponding part of the path integral in eq. (\ref{Idea:vev}) becomes
\begin{equation}
\int_{\varphi\gtrsim\varphi_0}\mathcal{D}\varphi\:\varphi(0)e^{-S}\rightarrow\varphi_0\int_{\bar{\varphi}\gtrsim 0}\mathcal{D}\bar{\varphi}Je^{-W} \; ,
\end{equation}
where 
\begin{equation}
W=-\bar{\varphi}(0)+S 
\end{equation}
and $J$ is a Jacobian of the transformation (\ref{Idea:NewVar}).

Next, we want to evaluate the vev (\ref{Idea:vev}) in the saddle-point approximation (SPA). The partition function is evaluated via a ground state configuration. Suppose that the functional $W$ admits appropriate saddle points through which the modified path integral can be evaluated as well. Then,
\begin{equation}\label{Idea:NewVevSPA}
\langle\varphi\rangle\sim \varphi_0 e^{-\bar{W}+S_0} \; .
\end{equation}
In this expression, $\bar{W}$ is the value of $W$ at a saddle and $S_0$ is the value of $S$ at the ground state.

Clearly, the possible saddles of the functional $W$ solve equations of motion for the field $\bar{\varphi}$ everywhere except the origin. At the origin, they satisfy the equation provided that the latter is supplemented with an instantaneous source of $\bar{\varphi}$,
\begin{equation}
\bar{\varphi}(0)=\int d^4x j(x)\bar{\varphi}(x) \; , ~~~~ j(x)=\delta^{(4)}(x) \; .
\end{equation}
The solutions of the equation with the source are expected to be singular at the point where the source acts. Despite this, they are valid saddle points of $W$ (but not $S$).

Let us discuss the conditions under which the transition from eq. (\ref{Idea:vev}) to eq. (\ref{Idea:NewVevSPA}) is possible. First, the theory must admit the singular configurations of the type described above, which approach the classical ground state away from the singular point. Second, the SPA must be justified by the presence in the theory of a suitable semiclassical parameter. The appearance of such parameter would ensure that $\bar{W}\gg 1$. If a particular calculation reveals $\bar{W}$ to be of the order of one or negative, one concludes that eq. (\ref{Idea:NewVevSPA}) is not valid. Last, but not least, a physical argumentation is necessary in order to justify the change of the field variable made in eq. (\ref{Idea:NewVar}). 

In section \ref{Sec:DD} we will see in detail how the conditions mentioned above are satisfied in a particular class of models comprising gravity and scalar fields. Here we just note that these conditions are, in fact, quire restrictive. It is easy to make sure that neither theories of a scalar field with no back-reaction on gravity nor theories with dynamical gravity and a minimal coupling of it to the scalar field possess classical configurations which would allow to arrive at eq. (\ref{Idea:NewVevSPA}).

Note that, because of the presence of gravity, the euclidean path integral in eq. (\ref{Idea:vev}) must be taken with caution. Indeed, it is known that the action of the euclidean quantum gravity is unbounded below; in particular, it suffers from the so-called conformal factor problem \cite{Gibbons:1977zz} (see also the discussion in \cite{Gratton:1999ya}). We assume that the properties of the theory in the UV regime result in a resolution of this problem in one or another way.

Eqs. (\ref{Idea:vev})---(\ref{Idea:NewVevSPA}) admit a straightforward generalization to the case when the vacuum geometry is not flat. In this case, the action of the theory must be supplemented by an appropriate boundary term, and the exponent in eq. (\ref{Idea:NewVevSPA}) will include the difference of the boundary terms taken at the ground state and at the configuration extremizing $W$. As will be shown in section \ref{Sec:D}, the presence of the cosmological constant is not relevant for the analysis of classical configurations whose characteristic scale $\varphi_0$ is associated with the Planck scale. Note also that the non-zero vacuum energy can be realized in a SI theory without an explicit breaking of the scale symmetry \cite{Shaposhnikov:2008xb, Blas:2011ac, Karananas:2016grc}.

The prefactor in eq. (\ref{Idea:NewVevSPA}) includes a parameter $\varphi_0$ with the dimension of mass. In SI theories a dimensionful parameter can arise due to a spontaneous symmetry breaking. If a theory possesses only one such parameter at the classical level, the vev of the field will inevitably be proportional to it. In this case, the quantity $\bar{W}$ can be viewed as a rate of suppression of the classical scale. Hence, eq. (\ref{Idea:NewVevSPA}) indicates the emergence of the hierarchy of scales, one of which is generated classically, and the other --- non-perturbatively. 

In evaluating the vev $\langle\varphi\rangle$ in the leading-order SPA, the fields of the theory, which do not participate in building the instanton configuration, are kept classically at their vacuum values. Fluctuations of the fields on top of the instanton are the source of perturbative corrections to the prefactor in eq. (\ref{Idea:NewVevSPA}). Evaluation of the prefactor with the accuracy beyond the naive dimensional analysis is difficult and is outside the scope of the present paper. However, the applicability of the SPA enables us to believe that the corrections coming with the fluctuation factor do not spoil the hierarchy of scales observed in the leading-order analysis. Moreover, as we will see in sections \ref{Sec:DD} and \ref{Sec:Pheno}, the instanton value of $W$ can vary depending on the parameters of the theory, and this can compensate possible deviations of the value of the vev, caused by subleading contributions.

\section{Dilaton model: exact euclidean configurations}
\label{Sec:D}

In this section, we consider a simple model admitting exactly solvable classical euclidean equations of motion. We will refer to it as the Dilaton model. We focus on the configurations solving these equations provided that the latter are accompanied with a scalar field source. These configurations share many important properties with their counterparts arising in more complicated theories of section \ref{Sec:DD}. The results of this section will provide us with an intuition about certain properties a theory must possess in order to permit the mechanism of generating the hierarchy of scales, which was outlined in section \ref{Sec:Idea}.

\subsection{The Lagrangian}

Consider the simplest SI model of one real scalar field coupled to gravity in a non-minimal way. The Lagrangian of the model is
\begin{equation}\label{D_L_J}
\dfrac{\L}{\sqrt{g}}=-\dfrac{1}{2}\xi\varphi^2 R+\dfrac{1}{2}(\d\varphi)^2+\dfrac{\lambda}{4}\varphi^4 \; ,
\end{equation}
where $(\d\varphi)^2\equiv g^{\mu\nu}\nabla_\mu\varphi\nabla_\nu\varphi$ and the non-minimal coupling constant $\xi$ is taken to be positive. The euclidean action of the model,
\begin{equation}\label{D_Action}
S=\int d^4x\L \; ,
\end{equation}
must be supplemented with an appropriate boundary term (see, e.g., \cite{Padilla:2012ze}). As we will see shortly, the latter should be taken in the form
\begin{equation}\label{D_BT}
I=-\int d^3x\sqrt{\gamma}K\xi\varphi^2 \; ,
\end{equation}
where $K$ denotes the external curvature of the space boundary and $\gamma$ the determinant of the metric induced on the boundary.

The model is invariant under the global scale transformations\footnote{The symmetry associated with the absence of dimensionful parameters can equivalently be written as an internal transformation, $g_{\mu\nu}(x)\mapsto q^{-2}g_{\mu\nu}(x)$, $\varphi(x)\mapsto q\varphi(x)$. }
\begin{equation}\label{D_ST_J}
g_{\mu\nu}(x)\mapsto g_{\mu\nu}(qx) \; , ~~~ \varphi(x)\mapsto q\varphi(q x)
\end{equation}
with $q$ a constant. Further, it admits the classical ground state of the form
\begin{equation}\label{D_CGS_J}
\varphi=\varphi_0 \; , ~~~ R=\dfrac{\lambda\varphi_0^2}{\xi} \; .
\end{equation}
The latter breaks scale symmetry spontaneously by introducing a classical scale $\varphi_0$.

To simplify the analysis of classical configurations, it is convenient to rewrite the model in the form in which the non-minimal coupling is absent. To this end, we perform a Weyl transformation of the metric field.\footnote{The scalar-tensor theories, related to each other by a Weyl rescaling of the metric, are classically equivalent. For the discussion of their equivalence at the quantum level see, e.g, \cite{Kamenshchik:2014waa}.} To keep the kinetic term of the scalar field canonical (up to a constant multiplier) in the new coordinates, we also redefine the scalar field variable:
\begin{equation}\label{D_Weyl}
\varphi=\varphi_0\Omega \; , ~~~ \tilde{g}_{\mu\nu}=\Omega^2g_{\mu\nu} \; , ~~~ \Omega=e^{\frac{\bar{\varphi}}{\sqrt{\xi}\varphi_0}} \; .
\end{equation}
The condition $\varphi>0$ implied by eq. (\ref{D_Weyl}) is not restrictive. In what follows, we will discuss the classical configurations which are monotonically-decreasing functions of a radial coordinate with the large-distance asymptotics $\varphi\rightarrow\varphi_0$.

The action becomes\footnote{Transformation of different quantities under the Weyl rescaling can be found, e.g., in \cite{Carneiro:2004rt}.}
\begin{equation}\label{L_transform}
S = \int d^4x\tilde{\L}-3\int d^4x\sqrt{\tilde{g}}\xi\varphi_0^2\tilde{\square}\log\Omega  \; ,
\end{equation}
where $\tilde{\square}\equiv\tilde{g}^{-1/2}\d_\mu\tilde{g}^{\mu\nu}\d_\nu$. The exterior curvature transforms as
\begin{equation}\label{K_transform}
K=\Omega\tilde{K}+3\tilde{n}^\mu\d_\mu\Omega \; ,
\end{equation}
where $\tilde{n}^\mu$ is a unit normal to the boundary in the coordinate frame provided by $\tilde{g}_{\mu\nu}$. After using Gauss's theorem, the second contribution in eq. (\ref{K_transform}) cancels the total derivative term in eq. (\ref{L_transform}). The transformed Lagrangian and boundary term are written as\footnote{Note that without the boundary term taken into account, the Lagrangian (\ref{D_L_E}) would contain the total derviative term, according to eq. (\ref{L_transform}), and this term would contribute to the action of a singular configuration, thus leading to an incorrect result.} 
\begin{equation}\label{D_L_E}
\dfrac{\tilde{\L}}{\sqrt{\tilde{g}}}=-\dfrac{1}{2}\xi\varphi_0^2\tilde{R}+\dfrac{1}{2a}(\tilde{\d}\bar{\varphi})^2+\dfrac{\lambda}{4}\varphi_0^4 \; , ~~~ a=\dfrac{1}{6+1/\xi} \; ,
\end{equation}
\begin{equation}\label{D_BT_E}
I_{GH}=-\xi\varphi_0^2\int d^3x\sqrt{\tilde{\gamma}}\tilde{K} \; .
\end{equation}
where we denote $(\tilde{\partial}\bar{\varphi})^2\equiv \tilde{g}^{\mu\nu}\nabla_\mu\bar{\varphi}\nabla_\nu\bar{\varphi}$. In the first term of the Lagrangian (\ref{D_L_E}) we recognize the Planck mass, 
\begin{equation}
M_P\equiv\sqrt{\xi}\varphi_0 \; ,
\end{equation}
and eq. (\ref{D_BT_E}) represents the usual Gibbons-Hawking term \cite{Gibbons:1976ue}, which justifies eq. (\ref{D_BT}).\footnote{Of course, one can check directly that the boundary term (\ref{D_BT}) cancels the surface terms arising from the variation of the metric in the action (\ref{D_Action}).} In the new coordinates, the scale transformations act as
\begin{equation}
\bar{\varphi}(x)\mapsto\bar{\varphi}(x)+q \; , ~~~ \tilde{g}_{\mu\nu}(x)\mapsto\tilde{g}_{\mu\nu}(x) \; .
\end{equation}

We see that the scalar field variable $\varphi$ is related to the canonical variable $\bar{\varphi}$ via the exponential mapping, according to eq. (\ref{D_Weyl}). Here one can see an analogy with the gauge theories, in the confinement phase of which the description must be performed in terms of the Wilson loops, not the gauge field itself \cite{Polyakov:1978vu}. Hence, the non-minimal coupling to gravity leads naturally to the appearance of the source term for $\bar{\varphi}$ in the process of evaluation of the vev $\langle\varphi\rangle$. In section \ref{Sec:DD}, this observation will enable us to write eq. (\ref{Idea:NewVevSPA}) for a classically zero vev of the scalar field.

\subsection{Classical configurations and Instanton action}
\label{Ssec:Class}

In studying classical configurations arising in the Dilaton model, we restrict ourselves to the spherically-symmetric case. This is motivated by the fact that introducing the instantaneous source of the scalar field does not break the $O(4)$-symmetry present in the theory. Should the less symmetric configurations suitable for our purposes exist, we assume that their contribution to the path integral is suppressed.\footnote{Although it was proven that the solution of maximal symmetry saturates the action in flat space background \cite{Coleman:1977th,Blum:2016ipp}, no such proof is known in the case when gravity dynamics is included.} Furthermore, below we neglect the curvature of the background solution (\ref{D_CGS_J}) by assuming that it has no impact on relevant properties of classical configurations whose characteristic energy scale exceeds significantly the scale of the background. This expectation is justified in appendix \ref{AppZ} where the case of non-zero $R$ is considered.

We adopt the following ansatz for the metric field,
\begin{equation}\label{Ansatz}
d\tilde{s}^2=f^2dr^2+r^2d\Omega_3^2 \; .
\end{equation}
Here $f$ is a function of the radial coordinate $r$ and $d\Omega_3$ is the line element of a unit 3-sphere. The scalar field equation of motion and 00-component of the Einstein equations read as follows,
\begin{equation}\label{D_EOM}
\d_r\left(\dfrac{r^3\bar{\varphi}'}{af}\right)=0 \; , ~~~ \dfrac{1}{f^2}=1+\dfrac{r^2\bar{\varphi}'^2}{6aM_P^2} \; .
\end{equation}
Note that thanks to the somewhat nonstandard form of the metric ansatz, the second of eqs. (\ref{D_EOM}) is an algebraic equation on $f$.

Equations of motion admit a solution of the form
\begin{equation}\label{D_CGS_E}
\bar{\varphi}=0 \; , ~~~ f=1 \; ,
\end{equation}
which represents the classical ground state (\ref{D_CGS_J}) of the model with $R=0$. To find other configurations, we replace the first of eqs. (\ref{D_EOM}) by
\begin{equation}\label{D_EOM_Mod}
\dfrac{r^3\bar{\varphi}'}{af}=C
\end{equation}
with $C$ some non-zero constant. We require the classical configuration obeying eq. (\ref{D_EOM_Mod}) to approach the vacuum solution (\ref{D_CGS_E}) at large distances. With this boundary condition, we obtain a one-parameter family of configurations distinguished by the value of $C$. Near the origin, the scalar field and the curvature behave as\footnote{The configurations of this type were studied before in the context of the cosmological initial value problem \cite{Hawking:1998bn,Garriga:1998tm,Vilenkin:1998pp,Turok:1999fe}. In those works, they are referred to as singular instantons. In this paper, we prefer to keep this name for a unique configuration of the family, satisfying an additional boundary condition, see eq. (\ref{D_EOM_Source}). } 
\begin{equation}\label{D_ShortAs}
\bar{\varphi}\sim -\gamma M_P\log(M_Pr) \; , ~~~ \tilde{R}\sim aM_P^{-4}r^{-6} \; , ~~~ \gamma=\sqrt{6a} \; , ~~~ r\rightarrow 0 \; .
\end{equation}
One observes that the physical singularity forms at the center of the configurations. Therefore, they are not valid solutions of eqs. (\ref{D_EOM}) at $r=0$. 

\begin{figure}[t]
\begin{center}
\center{\includegraphics[width=0.99\linewidth]{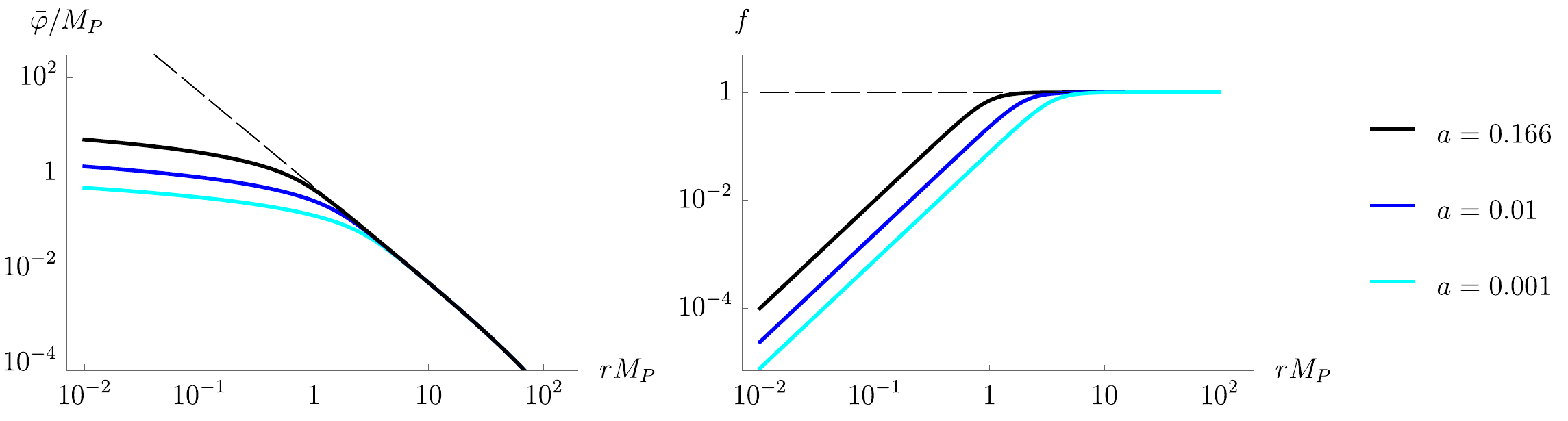}}
\caption{The profile of the singular instanton for the different values of $a$. The left panel demonstrates the logarithmic divergence of the scalar field $\bar{\varphi}$ in the core region of the instanton. The right panel shows the behavior of the metric function $f$ in the same region. The dashed lines represent the static gravity limit. }
\label{Fig:D_Sol}
\end{center}
\end{figure}

The divergence of a classical field configuration can be associated with a source of the field acting at the points of divergence. Therefore, such configuration can be regarded as a solution of equations of motion following from varying the action supplemented by a source term,
\begin{equation}\label{D:W}
W=S-\int d^4x j(x)\bar{\varphi}(x) \; .
\end{equation}
To reproduce the asymptotics (\ref{D_ShortAs}), the source $j(x)$ must be instantaneous,
\begin{equation}
j(x)=M_P^{-1}\delta^{(4)}(x) \; ,
\end{equation}
where we normalize the delta-function on the Planck scale as the latter is the only classical scale of the model. One of the singular configurations found above is obtained as a saddle point of the functional $W$. It is specified by
\begin{equation}\label{D_EOM_Source}
C=-M_P^{-1} .
\end{equation}
This can be viewed as an additional boundary condition fixing the position of the center of the singular configuration and the strength of the source producing it. In what follows, we will call the solution of eqs. (\ref{D_EOM_Mod}), (\ref{D_EOM_Source}) the singular instanton. It is explicitly given by
\begin{equation}\label{D_SingInst}
\bar{\varphi}(r)=\sqrt{\dfrac{3a}{8}}M_P\log\left[ \dfrac{\sqrt{1+6a^{-1}M_P^4r^4}+1}{\sqrt{1+6a^{-1}M_P^4r^4}-1} \right] \; , ~~~ \dfrac{1}{f^2(r)}=1+\dfrac{a}{6M_P^4r^4} \; .
\end{equation}
As is seen from this equation, the singular instanton has a characteristic length scale $a^{1/4}M_P^{-1}$ determining the size of its core. In the core region, the gravitational field is affected strongly by the dynamics of the scalar field. In turn, the short-distance behavior of the scalar field is affected by gravity, see figure \ref{Fig:D_Sol} for illustration.

We would like to note that the short-distance logarithmic divergence of the scalar field, expressed in eqs. (\ref{D_ShortAs}), reveals a nontrivial interplay between the scalar and gravitational sectors of the model. Indeed, in the flat space limit, the field $\bar{\varphi}$ in four dimensions exhibits the usual power-like massless asymptotics $\bar{\varphi}\sim r^{-2}$. We observe that gravity cures partially this divergence. This seems to be a promising sign of a general non-perturbative effect caused by gravity on the correlation functions in the scalar sector.\footnote{Note that the solution of eq. (\ref{D_EOM_Source}) can be viewed as an euclidean Green function of the massless scalar field propagating in the external gravitational background specified by the second of eqs. (\ref{D_SingInst}).}

Finally, we compute the euclidean action $\bar{S}$ and the boundary term $\bar{I}_{GH}$ of the singular instanton in the limit $\lambda=0$, relative to the action $S_0$ and the boundary term $I_{GH,0}$ of the background solution. This gives
\begin{equation}\label{Difference}
\bar{I}_{GH}-I_{GH,0}\sim a^{-1}M_P^{-2}r_s^{-2}\rightarrow 0 \; , ~~~ r_s\rightarrow\infty \; , ~~~ \lambda=0 \; ,
\end{equation}
where $r_s$ is the radius of a 3-sphere, and
\begin{equation}
\bar{S}=S_0=0 \; .
\end{equation}
Hence, there is no contribution of order $1$ from the instanton to the net euclidean action, neither to the net boundary term. Switching on the coupling $\lambda$ does not change this result, once the instanton and the cosmological scales are well separated, see appendix \ref{AppZ} for details.

To summarize, the singular instanton found above is a legitimate solution of the variational problem $\delta W/\delta\bar{\varphi}=0$ with $W$ given in eq. (\ref{D:W}). This goes in accordance with the logic presented in section \ref{Sec:Idea}. However, in the Dilaton model, the classical scale is defined by the vev of the scalar field, and there is no room for the second scale to be generated via the singular instanton. Moreover, as we just saw, the model is not capable of providing the large instanton action. To fix these drawbacks, one should change the structure of the model in the region of large $\bar{\varphi}$ and supplement it with the second scalar field whose vev is classically zero. We will proceed to this in section \ref{Sec:DD}.

\section{Instantons in scale-invariant gravity with two scalar fields}
\label{Sec:DD}

In this section, we study the class of SI models containing two scalar fields coupled to dynamical gravity. For convenience, the scalar fields are arranged into a two-component vector $\vec{\varphi}=(\varphi^1,\varphi^2)^T$. We are interested in the case when the vev of one of the fields is classically non-zero and can be associated with the Planck scale. By studying singular instantons similar to those of the Dilaton model of section \ref{Sec:D}, we will show how they can contribute to the vev of the second scalar field. As it will turn out, the new scale associated with this vev can, in fact, be many orders of magnitudes smaller than the Planck scale.

\subsection{Model setup}
\label{Ssec:ModelSetup}

In a search for a particular model in which one can realize quantitatively the reasoning outlined in section \ref{Sec:Idea}, we are guided by several principles. First, the model must be convertible into a phenomenologically viable theory upon identifying one of its scalar fields with the Higgs field dof and supplementing it with the rest of the SM content. Second, we require the model to enjoy global scale symmetry. This can ensure the stability of the scalar fields vev against perturbative quantum corrections, as will be discussed in detail in section \ref{Sec:Pheno}. Further, the results of studies of the Dilaton model give us certain hints about the desirable structure of the theory in the regime probed by the core of the singular instanton.

We choose the Lagrangian of the model in the following general form,
\begin{align}\label{DD_GenLagr_J}
\dfrac{\L}{\sqrt{g}}=-\dfrac{1}{2}\mathcal{G}(\vec{\varphi})R & +\dfrac{1}{2}\gamma^{(2)}_{ij}(\vec{\varphi})g^{\mu\nu}\d_\mu\varphi^i\d_\nu\varphi^j  \\ \nonumber
& +\sum_{n=2}^\infty\gamma^{(2n)}_{i_1,...,i_{2n}}(\vec{\varphi})g^{\mu\nu}\d_\mu\varphi^{i_1}\d_\nu\varphi^{i_2}...g^{\rho\sigma}\d_\rho\varphi^{i_{2n-1}}\d_\sigma\varphi^{i_{2n}} + V(\vec{\varphi}) \; .
\end{align}
The action of the model must be supplemented with the boundary term (cf. eq. (\ref{D_BT}))
\begin{equation}\label{DD_BD}
I=-\int d^3x\sqrt{\gamma}K\mathcal{G}(\vec{\varphi}) \; .
\end{equation}
The model is required to be invariant under the global scale transformations (cf. eq. (\ref{D_ST_J}))\footnote{For the sake of simplicity, we choose scaling dimensions of the scalar fields to be equal $1$.}
\begin{equation}\label{DD_ST}
g_{\mu\nu}(x)\mapsto g_{\mu\nu}(qx) \; , ~~~~ \vec{\varphi}(x)\mapsto q\vec{\varphi}(qx) \; .
\end{equation}
Next, we require the model to admit the classical ground state with the constant value of the Ricci scalar and 
\begin{equation}\label{DD_vac}
\vec{\varphi}_{vac.}=\left(\begin{array}{c}
\varphi_0 \\ 0
\end{array}\right) \; .
\end{equation}
Comparing with section \ref{Sec:D}, we see that $\varphi_1$ plays the role of the dilaton field. Finally, the derivative part of the Lagrangian must be organized so that to avoid the appearance of ghosts. We will specify the latter condition quantitatively when we rewrite the Lagrangian in the form which is more suitable for analytical analysis.

The functions introduced in the Lagrangian are taken as follows,\footnote{The indices of the components of the vector $\vec{\varphi}$ are raised and lowered with the euclidean metric $\delta_{ij}$ in the field space. }
\begin{align}
\mathcal{G} & =\xi_1\varphi_1^2+\xi_2\varphi_2^2 \; , \nonumber \\ 
\gamma^{(2)}_{ij} & =\delta_{ij}+\varkappa\mathcal{G}\mathcal{F}\mathcal{J}^{-4}(1+6\xi_i)(1+6\xi_j)\varphi_i\varphi_j \; , \nonumber \\
\gamma^{(4)}_{ijkl} & =\delta \mathcal{J}^{-8}(1+6\xi_i)(1+6\xi_j)(1+6\xi_k)(1+6\xi_l)\varphi_i\varphi_j\varphi_k\varphi_l \; , \label{DD_DefModel} \\ 
\gamma^{(2n)}_{i_1...i_{2n}} & =0 \; , ~~~~ n>2 \; . \nonumber
\end{align}
Here
\begin{equation}\label{DD_DefModelCurrent}
\mathcal{J}^2=(1+6\xi_1)\varphi_1^2+(1+6\xi_2)\varphi_2^2 \; ,
\end{equation}
\begin{equation}\label{DD_DefModelF}
\mathcal{F}=\dfrac{(1+6\xi_1)\varphi_2^2}{(1+6\xi_2)\varphi_1^2+(1+6\xi_1)\varphi_2^2} \; ,
\end{equation}
and $\xi_1$, $\xi_2$, $\varkappa$ and $\delta$ are constants. The potential for the scalar fields is chosen as
\begin{equation}
V=\dfrac{\lambda}{4}\varphi_2^4 \; . \label{DD_DefModelPot}
\end{equation}

The comments are in order on this choice of the ingredients of the model. The first of eqs. (\ref{DD_DefModel}) represents the simplest compatible with the symmetries non-minimal coupling of the scalar fields to gravity. It is of the same form as in the Dilaton model, in which it was shown to lead naturally to the appearance of the scalar field source when evaluating its vev.

The second of eqs. (\ref{DD_DefModel}) specifies the quadratic in derivatives part of the scalar sector of the model. The parameter $\varkappa$ controls its deviation from the canonical form. For the sake of simplicity, in sections \ref{Ssec:PolarVar}---\ref{Ssec:NonzeroDelta} we consider the case $\varkappa=0$, while the general case is postponed until section \ref{Ssec:NonzeroKappa}. There, we will see that $\varkappa$ serves to regulate certain properties of the singular instanton and instanton action near the source.

The third of eqs. (\ref{DD_DefModel}) determines the quartic in derivatives kinetic term of the model. It is absent in the Dilaton model, and, as we will see, it plays a crucial role in controlling the short-distance behavior of the instanton. The derivative terms of higher degrees are set to zero, because the effect they produce is analogous to the one of the quartic term. We address this question in some detail in appendix \ref{AppA}. 

Finally, the scalar field potential (\ref{DD_DefModelPot}) is chosen so as to be in accordance with a real-world theory in which $\varphi_2$ is to be identified with the Higgs field dof. For the same reason, the coupling constant $\lambda$ may be chosen to be field-dependent in a way that does not spoil the scale invariance of the model. This dependence would mimic the RG evolution of the Higgs self-coupling in a realistic setting.\footnote{The field-dependence of a normalization point in RG equations is essential in maintaining the scale invariance of the theory at the perturbative quantum level, see section \ref{Ssec:HD_Outline}. Also, in what follows we neglect the running of other constants, since it does not change the results qualitatively. } Note that we do not introduce the interaction terms $\propto\varphi_1^2\varphi_2^2$ and $\propto\varphi_1^4$ into the classical potential, although their presence is allowed by scale symmetry. In other words, we require the scalar sector of the model to respect the ``shift symmetry'' of the dilaton field $\varphi_1$. As will be discussed in section \ref{Sssec:HD_Corr} on a concrete example, the shift symmetry protects the mass of $\varphi_2$ from radiative corrections.

Evidently, with the choice of the operators given above, the model is invariant under the scale transformations (\ref{DD_ST}). Requiring the quadratic part of the kinetic terms for $\vec{\varphi}$ to be positive-definite puts a constraint on $\varkappa$, which will be specified below. The positive-definiteness of the derivative sector at high energies is ensured by setting $\delta>0$. We also require
\begin{equation}\label{CondOnXi}
\xi_2>\xi_1>0 \; .
\end{equation}
Last but not least, it is straightforward to see that eq. (\ref{DD_vac}) defines the classical ground state of the model, in which
\begin{equation}
\mathcal{G}(\vec{\varphi}_{vac.})=\xi_1\varphi_0^2\equiv M_P^2 \; .
\end{equation}

\subsection{Polar field variables}
\label{Ssec:PolarVar}

Let us rewrite the Lagrangian of the model in the form convenient for the analysis of classical configurations. To this end, one performs a Weyl rescaling of the metric, aimed at disentangling the Ricci scalar from the scalar dof, and a certain redefinition of the scalar field variables. The Weyl transformation reads as follows (cf. eq. (\ref{D_Weyl})),
\begin{equation}
\tilde{g}_{\mu\nu}=\Omega^2 g_{\mu\nu} \; , ~~~~  \Omega^2=\dfrac{\mathcal{G}(\vec{\varphi})}{\mathcal{G}(\vec{\varphi}_{vac.})} \; .
\end{equation}
The Lagrangian becomes
\begin{align}\label{DD_L_W}
\dfrac{\tilde{\L}}{\sqrt{\tilde{g}}}=-\dfrac{1}{2}M_P^2\tilde{R} & +\dfrac{1}{2}\tilde{\gamma}_{ij}^{(2)}(\vec{\varphi})\tilde{g}^{\mu\nu}\d_\mu\varphi^i\d_\nu\varphi^j \nonumber \\
& +\gamma^{(4)}_{ijkl}(\vec{\varphi})\tilde{g}^{\mu\nu}\tilde{g}^{\rho\sigma}\d_\mu\varphi^i\d_\nu\varphi^j\d_\rho\varphi^k\d_\sigma\varphi^l+\tilde{V}(\vec{\varphi}) \; ,
\end{align}
where
\begin{equation}
\tilde{\gamma}^{(2)}_{ij}=\Omega^{-2}\left(\delta_{ij}+\dfrac{3}{2}M_P^2\d_i\log\Omega^2\d_j\log\Omega^2\right) \; , ~~~~ \tilde{V}(\vec{\varphi})=V(\vec{\varphi})\Omega^{-4} \; ,
\end{equation}
and we made use of eqs. (\ref{DD_DefModel}) with $\varkappa=0$. The boundary term (\ref{DD_BD}) transforms into the usual Gibbons-Hawking term. According to the discussion in section \ref{Ssec:Class}, we can safely exclude it from consideration.

Following \cite{GarciaBellido:2011de}, we now look for a suitable redefinition of the scalar field variables. To trace the actual scalar dof in the Lagrangian (\ref{DD_L_W}), we would like to bring the quadratic in derivatives part of the kinetic term to a diagonal form. In the new variables $\vec{\chi}=\vec{\chi}(\vec{\varphi})$, let the latter take the form
\begin{equation}
\dfrac{1}{2}\bar{\gamma}^{(2)}_{nm}(\vec{\chi}(\vec{\varphi}))\tilde{g}^{\mu\nu}\d_\mu\chi^n\d_\nu\chi^m \; .
\end{equation}
Then, one demands that
\begin{equation}\label{Eq1}
\bar{\gamma}^{(2)}_{12}(\vec{\chi}(\vec{\varphi}))=0 \; ,
\end{equation}
which provides us with a first-order differential equation on the two components of the vector $\vec{\chi}$, thus leaving some freedom in the choice of new variables. It will prove to be useful to choose $\chi^1$, $\chi^2$ in such a way that the scale transformations (\ref{DD_ST}) leave one of the fields intact, while shifting another by a constant,
\begin{equation}\label{NewST}
\chi^1\mapsto\chi^1+q \; , ~~~~ \chi^2\mapsto\chi^2 \; .
\end{equation}
From eqs. (\ref{NewST}) one sees that $\chi^1$, $\chi^2$ are reminiscent of polar coordinates on a plane on which the scale transformations act by an isotropic dilation by a factor $q$. To find an equation $\vec{\chi}(\vec{\varphi})$ must satisfy in this case, we make use of the Noether current associated with the scale symmetry of the model. In view of eq. (\ref{DD_ST}), the latter is given by
\begin{equation}
\sqrt{\tilde{g}}J^\mu=\dfrac{\d\tilde{\L}}{\d\d_\mu\varphi^i}\varphi^i \; .
\end{equation}
For simplicity, let us put $\delta=0$ for the moment. Then, on the one hand,
\begin{align}\label{11}
\sqrt{\tilde{g}}J^\mu & =\tilde{g}^{\mu\nu}\varphi^i\tilde{\gamma}^{(2)}_{ij}(\vec{\varphi})\d_\nu\varphi^j \nonumber \\
& =M_P^2\tilde{g}^{\mu\nu}\dfrac{\d_\nu \mathcal{J}^2}{\mathcal{G}}
\end{align}
with $\mathcal{J}^2$ given in eq. (\ref{DD_DefModelCurrent}). On the other hand, when expressed in terms of the variables $\vec{\chi}$ satisfying eq. (\ref{NewST}), the current becomes
\begin{equation}\label{12}
\sqrt{\tilde{g}}J^\mu=M_P\tilde{g}^{\mu\nu}\bar{\gamma}^{(2)}_{11}(\vec{\chi}(\vec{\varphi}))\d_\nu\chi^1 \; .
\end{equation}
Equating (\ref{11}) and (\ref{12}), we obtain two more equations on $\vec{\chi}$. One can show that they are compatible and, combined with eq. (\ref{Eq1}), can be simultaneously solved. Denote this solution by $\chi^1=\rho$, $\chi^2=\theta$. Then, its explicit form is
\begin{equation}
\rho=\dfrac{M_P}{2}\log\dfrac{\mathcal{J}^2}{M_P^2} \; , ~~~~ \theta=\arctan\left(\sqrt{\dfrac{1+6\xi_1}{1+6\xi_2}}\dfrac{\varphi_2}{\varphi_1}\right) \; .
\end{equation}
It is now straightforward to derive the form of the Lagrangian in the new variables. It is given by
\begin{align}\label{DD_L_NewVar}
\dfrac{\tilde{\L}}{\sqrt{\tilde{g}}}=-\dfrac{1}{2}M_P^2\tilde{R} & +\dfrac{1}{2a(\theta)}(\tilde{\d}\rho)^2+\dfrac{b(\theta)}{2}(\tilde{\d}\theta)^2 \\ 
& + \delta\dfrac{(\tilde{\d}\rho)^4}{M_P^4}+\tilde{V}(\theta) \nonumber
\end{align}
with \cite{GarciaBellido:2011de}
\begin{align}\label{DD_L_NewVar2}
a(\theta)=a_0(\sin^2\theta+\zeta\cos^2\theta) \; , ~~~~ b(\theta)=\dfrac{M_P^2\zeta}{\xi_1}\dfrac{\tan^2\theta+\xi_1/\xi_2}{\cos^2\theta(\tan^2\theta+\zeta)^2} \; ,
\end{align}
\begin{equation}\label{DD_NewPot}
\tilde{V}(\theta)=\dfrac{\lambda M_P^4}{4\xi_2^2}\dfrac{1}{(1+\zeta \cot^2\theta)^2} \; ,
\end{equation}
and
\begin{align}\label{DD_aSI}
\zeta=\dfrac{(1+6\xi_2)\xi_1}{(1+6\xi_1)\xi_2} \; , ~~~~~ a_0=\dfrac{1}{6+1/\xi_2} \; .
\end{align}
First, we note that, due to the invariance of the model under the scale transformations (\ref{NewST}), the field $\rho$ enters the Lagrangian only through derivatives. As we will see, this makes its role analogous to that of the field $\bar{\varphi}$ in the Dilaton model. Second, the form of the quartic derivative term becomes strikingly simple in the new variables. Its suppression by $M_P$ is due to the classical vev which is now given by
\begin{equation}\label{DD_CGS_NewVar}
\rho_{vac.}=\dfrac{M_P}{2}\log\dfrac{1+6\xi_1}{\xi_1} \; , ~~~~ \theta_{vac.}=0 \; .
\end{equation}
Hence, the higher-dimensional derivative term determines the structure of the theory at high energies. Regarding the classical analysis, this term starts to be important in the limit of large derivatives of the $\rho$-component of the instanton and, hence, is expected to change the behavior of the latter in this limit.

As was already mentioned, the fields $\rho$ and $\theta$ can be thought of as polar coordinates on the plane spanned by $\sqrt{1+6\xi_1}\varphi_1$ and $\sqrt{1+6\xi_2}\varphi_2$. In particular, $\theta$ is analogous to the angle on that plane, and $\rho$ --- to the logarithm of the radius. Because of this, in what follows we will refer to $\rho$ as the radial and to $\theta$ as the angular field variables.

Let us finally quote the inverse formulas,
\begin{equation}\label{DD_Inv}
\varphi_1=\dfrac{M_P\cos\theta}{\sqrt{1+6\xi_1}}\; e^{\rho/M_P} \; , ~~~~~ \varphi_2=\dfrac{M_P\sin\theta}{\sqrt{1+6\xi_2}}\; e^{\rho/M_P} \; . 
\end{equation}
One observes that the original scalar fields are expressed through the exponent of the field $\rho$. Hence, according to the discussion in section \ref{Sec:Idea}, the source of the radial field naturally appears in the course of evaluation of the vev of $\varphi_2$.\footnote{Although the change of variables (\ref{DD_Inv}) is applicable for all $\vec{\varphi}\neq\vec{0}$, one can think of $\rho$, $\theta$ as replacing the original scalar dof in the regime where the latter are not canonical, $\vert\varphi_1-\varphi_0\vert\gtrsim\varphi_0$, $\vert\varphi_2\vert\gtrsim\varphi_0$. } This points again at the similarity between $\rho$ and the field $\bar{\varphi}$ of the Dilaton model. 

Note also that from eq. (\ref{DD_GenLagr_J}) and the first of eqs. (\ref{DD_DefModel}) it follows that in the limit when $\varphi_1$ and $\varphi_2$ vanish simultaneously the model is not well-defined. The classical configurations we study below avoid this point; in fact, for them
\begin{equation}
\rho\geqslant\rho_{vac.} \; .
\end{equation}

\subsection{Instanton in a model without higher-dimensional terms}
\label{Ssec:ZeroDelta}

We begin to study classical configurations arising in the model specified by eqs. (\ref{DD_L_NewVar})---(\ref{DD_aSI}). We restrict ourselves to the analysis of $O(4)$-symmetric configurations and choose the metric Ansatz as in eq. (\ref{Ansatz}). The configuration must approach the classical ground state (\ref{DD_CGS_NewVar}) at infinity. Since the quartic derivative term affects only the short-distance part of the instanton, it is convenient to study first the case when $\delta=0$.

From eq. (\ref{DD_L_NewVar}) we obtain the equation of motion for the radial field $\rho$,
\begin{equation}\label{EqOnRho}
\d_r\left(\dfrac{\rho' r^3}{a(\theta)f}\right)=0 \; ,
\end{equation}
which is fully analogous to eq. (\ref{D_EOM}). Thanks to the form of the metric Ansatz and the fact that $\rho$ enters the Lagrangian derivatively, both $\rho'$ and $f$ can be expressed explicitly through the angular field $\theta$ and its derivatives. Therefore, finding a solution reduces to solving a single second-order differential equation on $\theta$. Switching on the source of $\rho$ selects a unique solution from the family of configurations obeying eq. (\ref{EqOnRho}). In view of eqs. (\ref{DD_Inv}), we specify the source as follows,
\begin{equation}\label{DD_W}
W=S-\int d^4x \delta^{(4)}(x)\rho(x)/M_P \; .
\end{equation}
Equation of motion becomes
\begin{equation}\label{EqOnRho2}
\dfrac{\rho' r^3}{f}=-\dfrac{a(\theta)}{M_P} \; .
\end{equation}
Let us focus on the classical configurations satisfying eq. (\ref{EqOnRho2}) and approaching the ground state (\ref{DD_CGS_NewVar}) at infinity. The large-distance asymptotics of these solutions are inferred directly from equations of motion, they coincide with the ones of the massless fields,\footnote{Note that self-consistency dictates the fields to approach the values corresponding to the actual vev of $\varphi_1$, $\varphi_2$. The difference can be neglected on practice provided that the characteristic size of the configuration contributing to the vev is much smaller than $\langle\varphi_2\rangle^{-1}$. }
\begin{equation}\label{LargeDistAs}
\rho-\rho_0\sim r^{-2} \; ,  ~~~ \theta\sim r^{-2} \; , ~~~ r\rightarrow\infty \; .
\end{equation}
We now turn to the short-distance behavior of the solutions. We require the fields constituting the instanton to behave monotonically with the distance. Then, the angular field must have a definite limit $\theta\rightarrow\theta_0$ at $r\rightarrow 0$. Inspecting eq. (\ref{EqOnRho2}) and $00$-component of the Einstein equations reveals that
\begin{equation}\label{AsOnRhoG}
\rho\sim -\gamma M_P\log M_Pr \; , ~~~\tilde{R}\sim r^{-6} \; , ~~~r\rightarrow 0 \; ,
\end{equation}
where $\tilde{R}$ is the Ricci scalar and
\begin{equation}\label{Gamma}
\gamma=\sqrt{6a_0} \; .
\end{equation}
We conclude that $\rho$ carries the same properties as the scalar field $\bar{\varphi}$ in the Dilaton model.

In looking for allowable values of $\theta_0$, we find it important to note that the values different from $\pi k/2$, $k=0,1,2,...$, are possible only if one requires \cite{Shkerin:2016ssc}
\begin{equation}\label{CondOnBounce}
\rho'=0 \; .
\end{equation}
A classical configuration with this property exists if the potential $\tilde{V}(\theta)$ becomes negative for some $\theta$. It is the bounce interpolating between the regions of false and true vacua \cite{Coleman:1977py,Coleman:1980aw}.\footnote{In \cite{Shkerin:2016ssc}, the bounce was studied in the context of EW vacuum stability in the Higgs-Dilaton model.} The solution we are interested in violates the condition (\ref{CondOnBounce}), hence it differs qualitatively from the possible bounce. One can show that the only admissible values of $\theta_0$ for this solution are
\begin{equation}\label{AsValueOfTheta}
\theta_0=\dfrac{\pi}{2}+\pi k \; , ~~~~ k=1,2,...
\end{equation}
We focus on the case $k=0$, since, as will be seen shortly, this is the only case when the configuration approaching the ground state at infinity exists. Then, one has
\begin{equation}
a(\theta_0)=a_0 \; .
\end{equation}
Recall that $a(\theta_0)$ regulates the strength of the source felt by the radial field. The short-distance asymptotics of $\theta$ is found to be
\begin{equation}\label{AsOnTheta}
\dfrac{\pi}{2}-\theta\sim r^\eta \; , ~~~~ r\rightarrow 0
\end{equation}
with
\begin{equation}\label{Beta}
\eta=\sqrt{6a_0(1-\xi_1/\xi_2)} \; ,
\end{equation}
provided that inequality (\ref{CondOnXi}) holds. The exponents (\ref{Gamma}) and (\ref{Beta}) demonstrate essential non-analyticity of the configuration in the core region, caused by the presence of the source. We will refer to the solution satisfying eqs. (\ref{LargeDistAs}), (\ref{AsOnRhoG}) and (\ref{AsOnTheta}) as the singular instanton.

To understand better the properties of the singular instanton near the source, we write its asymptotics in terms of the original field variables,
\begin{equation}\label{AsOldVar}
\varphi_1\sim r^{-\gamma+\eta} \; , ~~~\varphi_2\sim r^{-\gamma} \; . 
\end{equation}
Since $\eta<\gamma$, we conclude that both fields diverge at the center of the instanton. It is important to note that the divergence of $\varphi_1$, $\varphi_2$ originates fully from the divergence of the radial field. Hence, eqs. (\ref{DD_Inv}) provide a splitting of the scalar fields on the singular exponential part and the finite angular prefactor. The core region of the instanton is determined by the relation $\vert\varphi_2\vert\gg\vert\varphi_1\vert$ or, equivalently, $\vert\d\varphi_2\vert\gg\vert\d\varphi_1\vert$.

\begin{figure}[t]
\begin{center}
\center{\includegraphics[width=0.55\linewidth]{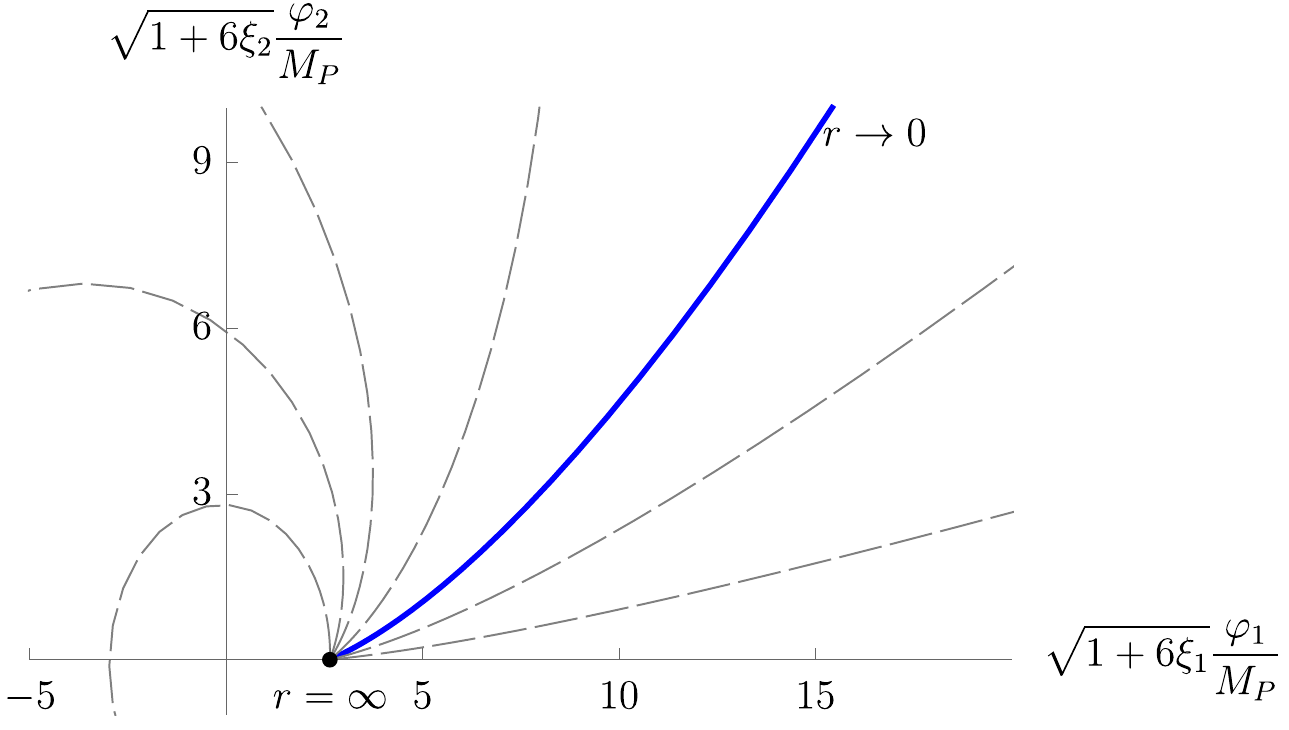}}
\caption{The classical configurations of the model (\ref{DD_L_NewVar})---(\ref{DD_aSI}), satisfying the vacuum boundary conditions, eq. (\ref{LargeDistAs}), at infinity. The solid blue line represents the singular instanton obeying eqs. (\ref{AsOnRhoG}), (\ref{AsOnTheta}). It is the only configuration with the finite asymptotics of $\theta$. The dashed lines show examples of other configurations. All solutions are distinguished by their fall-off at infinity, $\theta\sim cr^{-2}$, $r\rightarrow\infty$. The parameter $c$ is used as a shooting parameter in numerical calculations. The parameters of the model are $\xi_1=1$, $\xi_2=1.1$ and $\delta=\lambda=0$. }
\label{Fig:DD_Sol}
\end{center}
\end{figure}

As an example, figure \ref{Fig:DD_Sol} shows the singular instanton for a particular choice of parameters of the model. The solution is found by solving numerically equation for $\theta$, by the means of shooting. For illustrative purposes, the configurations with no limit of $\theta$ at $r\rightarrow 0$ are also shown. One can see from the figure that only for $\theta_0=\pi/2$ does the solution have the appropriate large-distance behavior.

The singular instanton of the type found above exists regardless the shape of the potential for the field $\theta$, encoded in the function $\lambda=\lambda(\theta)$. It is so because the potential does not affect neither long-distance nor short-distance asymptotics of the solution.\footnote{Of course, it does affect the solution in between two regions and, in particular, the value of the shooting parameter.} Note also that figure \ref{Fig:DD_Sol} demonstrates the difference of the singular instanton from the possible bounce, which is noticeable even in the limit $r\rightarrow\infty$. Indeed, from eqs. (\ref{DD_Inv}) and (\ref{CondOnBounce}) we see that for the bounce in this limit $d\varphi_2/d\varphi_1\rightarrow\infty$, while for the instanton the ratio remains finite.

\subsection{Regularization of the instanton by a higher-dimensional term}
\label{Ssec:NonzeroDelta}

Let us now switch on the Planck-suppressed quartic derivative operator in the Lagrangian (\ref{DD_L_NewVar}). It gives us a new ingredient, as compared to the Dilaton model of section \ref{Sec:D}. Importantly, the variation of this operator with respect to $\rho$ is a total derivative, hence, equation of motion for $\rho$ following from varying the functional (\ref{DD_W}) takes the form
\begin{equation}\label{DD_EOM2}
\dfrac{4\delta}{M_P^4}\dfrac{\rho'^3r^3}{f^3}+\dfrac{\rho'r^3}{a(\theta)f}=-\dfrac{1}{M_P} \; .
\end{equation}
This is again an exact equation. Denote by $\bar{r}$ the size of the region where the first term in eq. (\ref{DD_EOM2}) is dominant. In what follows, we will choose $\delta$ to be such that the length $\bar{r}$ is well within the region where $a(\theta)$ does not differ noticeably from its asymptotic value $a_0$. This will allow us to neglect the dynamics of the angular field when discussing the effects of the higher-dimensional operator on the behavior of the instanton. Note also that the interpretation of $a_0$ as the parameter regulating the strength of the source is preserved, since the short-distance asymptotics defined by eq. (\ref{EqOnRho2}) develops before the first term in eq. (\ref{DD_EOM2}) comes into play. 

\begin{figure}[t]
\begin{center}
\center{\includegraphics[width=0.55\linewidth]{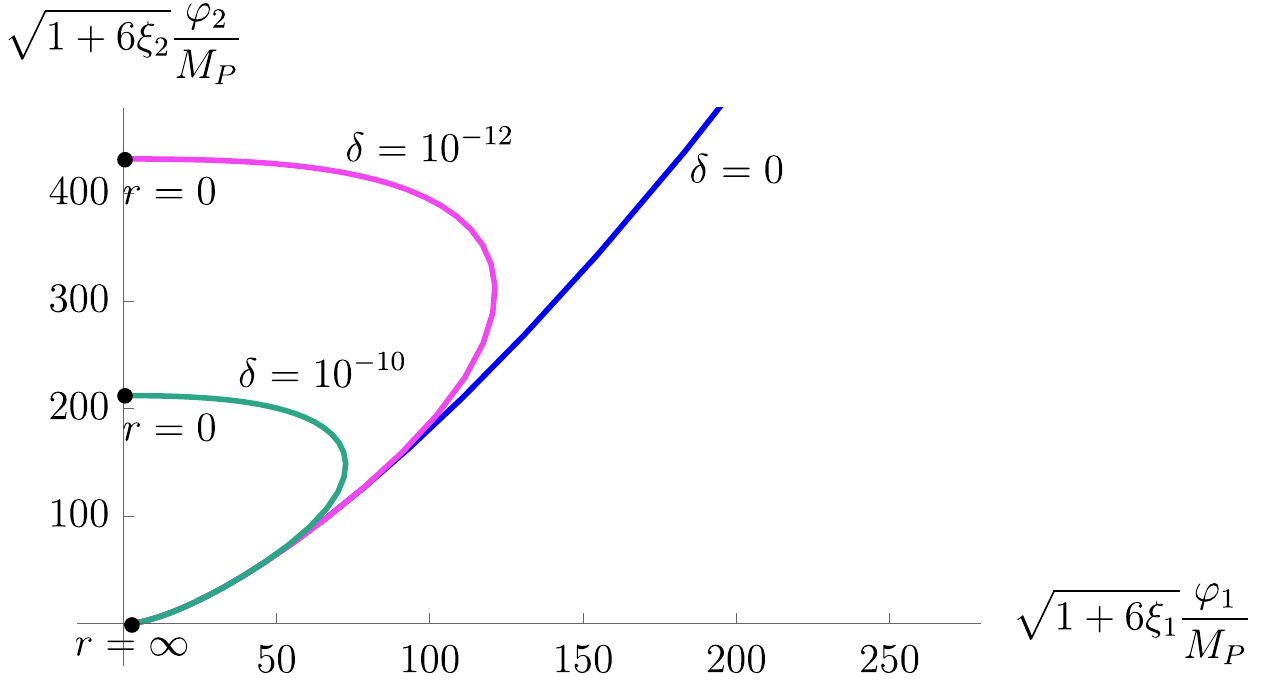}}
\caption{The family of singular instantons of the model (\ref{DD_L_NewVar})---(\ref{DD_aSI}), corresponding to different values of the parameter $\delta$ in the higher-dimensional derivative term. One observes that this derivative term regularizes the logarithmic divergence of the radial field and makes the latter finite at the center of the instanton. The parameters of the model are $\xi_1=1$, $\xi_2=1.1$ and $\lambda=0$.  }
\label{Fig:DD_SolReg}
\end{center}
\end{figure}

At $r\lesssim\bar{r}$, the behavior of the singular instanton is
\begin{equation}\label{NewAs}
\rho'\sim-M_P^2\delta^{-1/6} \; , ~~~ f\sim M_P r\delta^{1/6} \; .
\end{equation}
From this and eqs. (\ref{DD_EOM2}) and (\ref{AsOnRhoG}) one can infer the value of $\bar{r}$,
\begin{equation}\label{Size}
\bar{r}\sim M_P^{-1}\delta^{1/6}a_0^{1/2} \; .
\end{equation}
The crucial observation is that, thanks to the first of eqs. (\ref{NewAs}), the radial field is not divergent any more, and its magnitude at the center of the instanton is finite. It can be estimated from eqs. (\ref{AsOnRhoG}), (\ref{NewAs}) and (\ref{Size}) that
\begin{equation}\label{DD_Rho(0)}
\rho(0)/M_P\sim a_0^{1/2}(\log\delta-3\log a_0+\mathcal{O}(1)) \; .
\end{equation}
Despite the finiteness, the instanton remains to be singular. In particular, the scalar curvature behaves as (cf. eq. (\ref{AsOnRhoG}))
\begin{equation}
\tilde{R}\sim r^{-2} \; . 
\end{equation}
Therefore, introducing the source of the radial field is still a necessary step in obtaining the solution.

An example of how the higher-dimensional term regularizes the divergence of the instanton is presented in figure \ref{Fig:DD_SolReg}. Because of eqs. (\ref{NewAs}), the small values of $\delta$ are required to ensure the separation of the region where $a(\theta)$ varies from the region where the regularization acts. Note, however, that the smallness of $\delta$ does not bring in the model any new interaction scales below the Planck scale.

\subsection{Source enhancement}
\label{Ssec:NonzeroKappa}

\begin{figure}[t]
\begin{center}
\center{\includegraphics[width=0.85\linewidth]{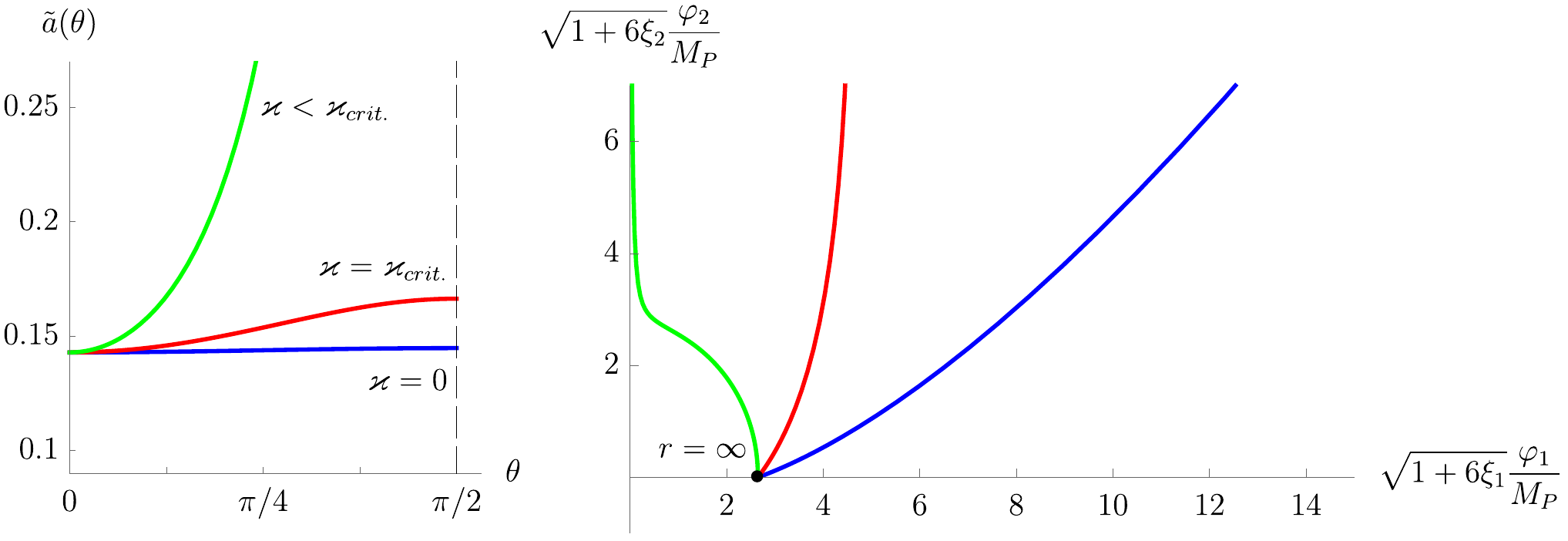}}
\caption{The singular instantons of the model (\ref{DD_L_NewVar})---(\ref{DD_aSI}) with $a(\theta)$ replaced by $\tilde{a}(\theta)$ according to eq. (\ref{DD_aMod}), and with $\varkappa$ varied. The left panel shows the function $\tilde{a}(\theta)$. In the limit $\varkappa=0$, the original model is reproduced. The critical value, $\varkappa=\varkappa_{crit.}$, corresponds to the case when $\eta=\gamma$ in eqs. (\ref{AsOldVar}), see appendix \ref{AppB} for details. The value below the critical, $\varkappa<\varkappa_{crit.}$, is chosen so that $\tilde{a}(\theta_0)\equiv\tilde{a}_0=100$. This value lies close to the positivity bound in eq. (\ref{CondOnKappa}). The right panel shows the corresponding instanton solutions. At $\varkappa=0$, the instanton studied in figures \ref{Fig:DD_Sol} and \ref{Fig:DD_SolReg} is reproduced. The parameters of the model are $\xi_1=1$, $\xi_2=1.1$ and $\lambda=0$. }
\label{Fig:DD_SolSource}
\end{center}
\end{figure}

From eq. (\ref{DD_Rho(0)}) one sees that the parameter $a_0=a(\theta_0)$, alongside with $\delta$, controls the large-$\rho$ properties of the singular instanton. In the model (\ref{DD_L_NewVar})---(\ref{DD_aSI}), the value of $a_0$ is determined by the non-minimal coupling $\xi_2$ and, according to eq. (\ref{DD_aSI}), is confined in the region
\begin{equation}\label{CondOnA}
0<a_0<1/6 \; .
\end{equation}

Since $a_0$ is associated with the strength of the source of the radial field, it is important to investigate the possibility that it can take values other than those prescribed by inequality (\ref{CondOnA}). In particular, we are interested in making the upper bound in this inequality arbitrarily large. This can be achieved by switching on the parameter $\varkappa$ in eqs. (\ref{DD_DefModel}), which was set to zero in the previous analysis. Starting from the Lagrangian in the form (\ref{DD_GenLagr_J}), we follow the steps performed in section \ref{Ssec:PolarVar} to obtain the description of the modified model in terms of the polar field variables. It is straightforward to see that the modified Lagrangian is still given by eq. (\ref{DD_L_NewVar}), but with the function $a(\theta)$ replaced by a new function $\tilde{a}(\theta)$ so that
\begin{equation}\label{DD_aMod}
\dfrac{1}{\tilde{a}(\theta)}=\dfrac{1}{a(\theta)}+\varkappa\sin^2\theta \; .
\end{equation}
As $\theta$ approaches the vacuum value, $\tilde{a}(\theta)$ becomes indistinguishable from $a(\theta)$, hence the properties of the model near the ground state remain unchanged. In particular, the large-distance properties of the singular instanton are independent of $\varkappa$.

Let us focus on the short-distance behavior of the instanton solution. One can make sure that the asymptotic value of $\theta$ obeys eq. (\ref{AsValueOfTheta}) with $k=0$ regardless the presence of $\varkappa$. Requiring the quadratic in derivatives part of the Lagrangian to be positive-definite yields
\begin{equation}\label{CondOnKappa}
\varkappa>-\dfrac{1}{a_0} \; .
\end{equation}
Varying $\varkappa$ within this region, one can achieve any positive strength of the radial field source $\tilde{a}_0\equiv\tilde{a}(\theta_0)$. 

In figure \ref{Fig:DD_SolSource} some particular values of $\varkappa$ are considered. One observes that the properties of the singular instanton at short distances depend significantly on the choice of $\varkappa$. The dependence is encoded in the exponents $\gamma$ and $\eta$ whose form for $\varkappa\neq 0$ is not given by eqs. (\ref{AsOnRhoG}), (\ref{AsOnTheta}) any more. Leaving the quantitative analysis to appendix \ref{AppB}, here we just note that $\eta$ exceeds $\gamma$ for $\varkappa$ lying close to the bound specified by eq. (\ref{CondOnKappa}). From eq. (\ref{AsOldVar}) we see that in this case the field $\varphi_1$ tends to zero as the source is approached even without the regularization provided by the quartic derivative term. The latter, however, is still necessary to remove the divergence of the field $\varphi_2$.

We would like to stress that the explicit form of the function $\tilde{a}(\theta)$ resulting in a particular source strength $\tilde{a}_0$ is, in fact, a matter of convenience, provided that the properties of the model near the ground state are respected. We choose this function according to eq. (\ref{DD_aMod}) because of the simple form it takes in the polar field variables and because it will fit well into the phenomenological analysis of section \ref{Sec:Pheno}. Finally, the effect produced by the quartic derivative term remains unchanged as long as $\tilde{a}(\theta)$ approaches the asymptotic value before this term takes over.

To summarize, in sections \ref{Ssec:ZeroDelta}---\ref{Ssec:NonzeroKappa} we have constructed and studied the singular instantons arising in the class of SI models specified by eqs. (\ref{DD_GenLagr_J}) and (\ref{DD_DefModel})---(\ref{DD_DefModelPot}). The principal difference of these models from the one-field Dilaton theory of section \ref{Sec:D} is the presence of two parameters, $\varkappa$ and $\delta$, associated with the structure of the theory at high energies, which determine its properties in the regime when $\vert\varphi_2\vert\gg\vert\varphi_1\vert$ and $\vert\d\varphi_2\vert\gg\vert\d\varphi_1\vert$. Namely, the parameter $\delta$ serves to regularize the logarithmic divergence of the radial field and to make $\rho(0)/M_P$ finite. As for $\varkappa$, its crucial role will be uncovered in the next section.

\subsection{New scale via the instanton}
\label{Ssec:NewScale}

As was already discussed, the ground state (\ref{DD_vac}) provides us with a single mass parameter $\varphi_0\approx M_P$ at the classical level, at least when the non-minimal couplings $\xi_1$, $\xi_2$ are of the order of one. We would like to see if the singular instantons obtained before can generate a new scale, by contributing non-perturbatively to the vev of $\varphi_2$. We are interested in the case when the contribution is such that the hierarchy 
\begin{equation}\label{TheHierarchy}
\langle\varphi_2\rangle/\langle\varphi_1\rangle\ll 1
\end{equation}
emerges. 
%Here we leave aside the question of perturbative corrections to the vev of the scalar fields. The assumption that the latter are not capable of generating the desired ratio in the l.h.s. of eq. (\ref{TheHierarchy}) will be justified in section \ref{Ssec:HD_Outline} in the context of the Higgs-Dilaton theory. For this purpose, we will employ the global scale symmetry of the theory and the requirement of the absence of heavy particle's thresholds above the EW scale.

Following the reasoning of section \ref{Sec:Idea}, we attempt to evaluate the vev of $\varphi_2$ with the new functional $W$. The latter is defined in eq. (\ref{DD_W}). The appropriate saddle points of $W$ are the singular instantons studied above. We will investigate if it is possible to adjust the parameters of the model to yield
\begin{equation}\label{CondOnW}
\bar{W}\gg 1\; ,
\end{equation}
where $\bar{W}$ is the instanton value of $W$. Applying the SPA, one arrives at
\begin{equation}\label{SPAScales}
\langle\varphi_2\rangle\sim M_P e^{-\bar{W}} \; .
\end{equation}
If for a particular choice of the parameters the condition (\ref{CondOnW}) is violated, one concludes that the SPA is not applicable and eq. (\ref{SPAScales}) is not valid. The possible interpretation of this case is that non-perturbative quantum gravity effects are strong and drive the value of $\langle\varphi_2\rangle$ close to $M_P$ so that no new scale appears. If, on the other hand, eq. (\ref{CondOnW}) is satisfied, these effects are suppressed, and the hierarchy of scales (\ref{TheHierarchy}) is generated. Note that the Planck mass appears as a prefactor in eq. (\ref{SPAScales}), as it is the only classical scale of the model.

\begin{figure}[t]
\begin{center}
\center{\includegraphics[width=0.99\linewidth]{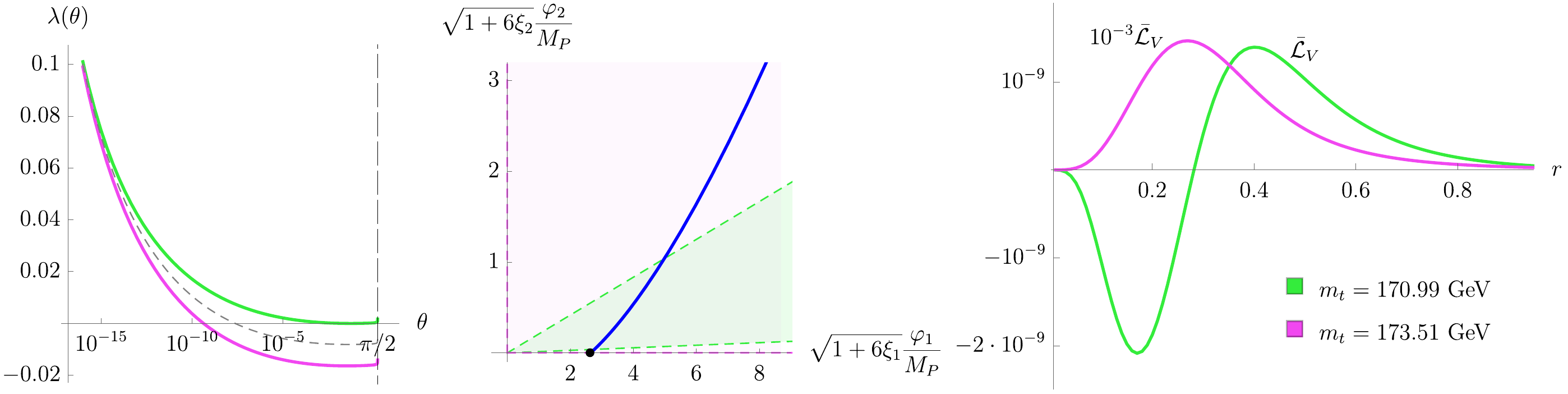}}
\caption{Left: the SM Higgs self-coupling $\lambda(\hat{\mu})$ at NNLO with the $\theta$-dependent momentum scale $\hat{\mu}$ given in eq. (\ref{DD_NormPoint}). The RG equations are solved using the code based on \cite{Chetyrkin:2012rz,Bezrukov:2012sa}. The solid lines represent the $2\sigma$-uncertainty region of the top quark mass, the dashed line corresponds to the central value $m_t=172.25$ GeV \cite{Castro:2017yxe}. The Higgs mass is taken to be $m_H=125.09$ GeV \cite{Aad:2015zhl}. Middle: the singular instanton in the potential (\ref{DD_NewPot}) with $\lambda$ plotted on the left side. The dashed lines encompass the regions of negative $\lambda$. One observes no difference between the solutions corresponding to the different choices of $\lambda$. Right: the potential part of the instanton Lagrangian, see eq. (\ref{LdLV}). One sees the contribution from $\bar{\L}_V$ to the instanton action $\bar{S}$ to be negligible compared to the overall contribution which is supposed to give $\bar{S}\gg 1$. The parameters of the model are $\xi_1=1$, $\xi_2=1.1$. }
\label{Fig:DD_Pot}
\end{center}
\end{figure}

Let us proceed to computation of $\bar{W}$. Since the potential $\tilde{V}$, given in eq. (\ref{DD_NewPot}), tends to zero when $\theta$ approaches its vacuum value, the geometry of the solution is asymptotically flat and the ground state action is zero. Contributions to $\bar{W}$ come from the source term and the instanton action $\bar{S}$. Making use of the Einstein equations and applying the Ansatz (\ref{Ansatz}), we have
\begin{equation}\label{WOnShell}
\bar{W}=-\dfrac{\rho(0)}{M_P}+\int_0^\infty dr (\bar{\L}_\delta-\bar{\L}_V) \; ,
\end{equation}
where
\begin{equation}\label{LdLV}
\bar{\L}_\delta=2\pi^2 r^3f\left(\dfrac{\rho'}{M_Pf}\right)^4 \; , ~~~ \bar{\L}_V=2\pi^2r^3f\tilde{V}(\theta) \; .
\end{equation}

We will study separately the contributions from the long-distance and short-distance parts of the instanton. The dominant term in the long-distance region is the one provided by the potential, $\bar{\L}_V$. According to eq. (\ref{DD_NewPot}), it is mainly determined by the quartic coupling $\lambda$. Bearing in mind phenomenological applications of our analysis, we consider $\lambda$ as a function of $\theta$ in order to mimic the RG evolution of the Higgs self-coupling in the SM setting.\footnote{The dependence of the self-coupling on the radial field would be inconsistent with the (quantum perturbative) scale invariance of the theory. } Specifically, we take the running of $\lambda$ corresponding to the $2\sigma$-uncertainty region around the central value of the top quark mass $m_t=172.25$ GeV \cite{Castro:2017yxe}, and to the central value of the Higgs mass $m_H=125.09$ GeV \cite{Aad:2015zhl}. The field-dependent momentum scale $\hat{\mu}=\hat{\mu}(\theta)$ is chosen according to the prescription (see section \ref{Ssec:HD_GenVEV})
\begin{equation}\label{DD_NormPoint}
\hat{\mu}^2=\dfrac{y_t^2}{2\xi_2}\dfrac{1}{1+\zeta\cot^2\theta} \; ,
\end{equation}
where $y_t$ is the top quark Yukawa coupling and $\zeta$ is given in eq. (\ref{DD_aSI}). With the potential specified in this way, we find the singular instanton numerically and compute its contribution to the potential part of the Lagrangian $\bar{\L}_V$. The results of the computation are shown in figure \ref{Fig:DD_Pot}. The main observation is that the potential term contributes negligibly to the instanton action. The reason lies in the fact that the instanton shoots too fast through the region where the action can be saturated by $\bar{\L}_V$. Hence, provided that we are interested in the total contribution to satisfy inequality (\ref{CondOnW}), one can safely ignore the potential term in eq. (\ref{WOnShell}). Note that this result points again at the qualitative difference between the singular instanton and the bounce for which the overall contribution comes exclusively from the potential. 

\begin{figure}[t]
\begin{center}
\center{\includegraphics[width=0.75\linewidth]{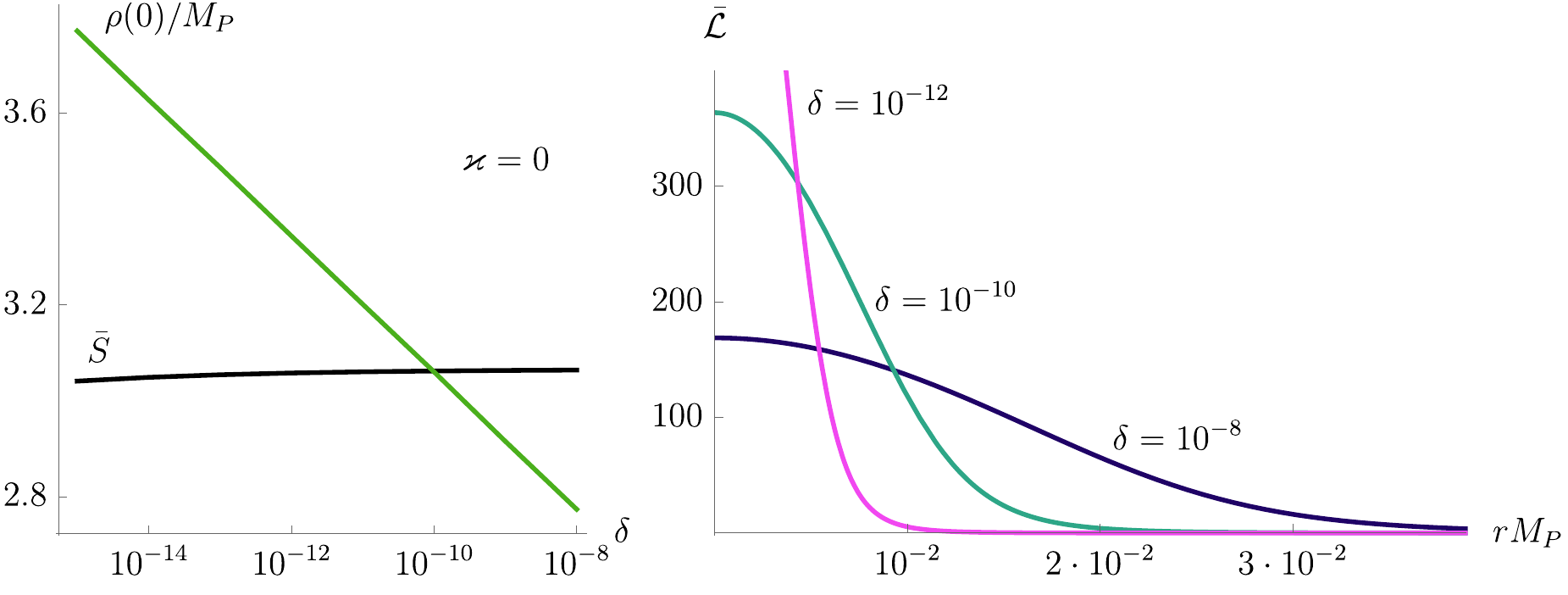}}
\caption{The singular term $\rho(0)/M_P$ and the instanton action $\bar{S}=\int dr\bar{\L}$, contributing to $\bar{W}$ according to eq. (\ref{WOnShell}). Here we take $\varkappa=0$ and $\xi_1=1$, $\xi_2=1.1$. The left panel shows the two contributions depending on the choice of $\delta$. One sees that, although $\bar{W}$ is positive for $\delta\gtrsim 10^{-10}$, it is impossible to achieve the regime when $\bar{W}\gg 1$. The right panel shows the instanton Lagrangian as a function of the radial coordinate and for different values of $\delta$. An agreement with eqs. (\ref{Size}) and (\ref{Ld}) is observed. }
\label{Fig:DD_Delta}
\end{center}
\end{figure}

\begin{figure}[b]
\begin{center}
\center{\includegraphics[width=0.6\linewidth]{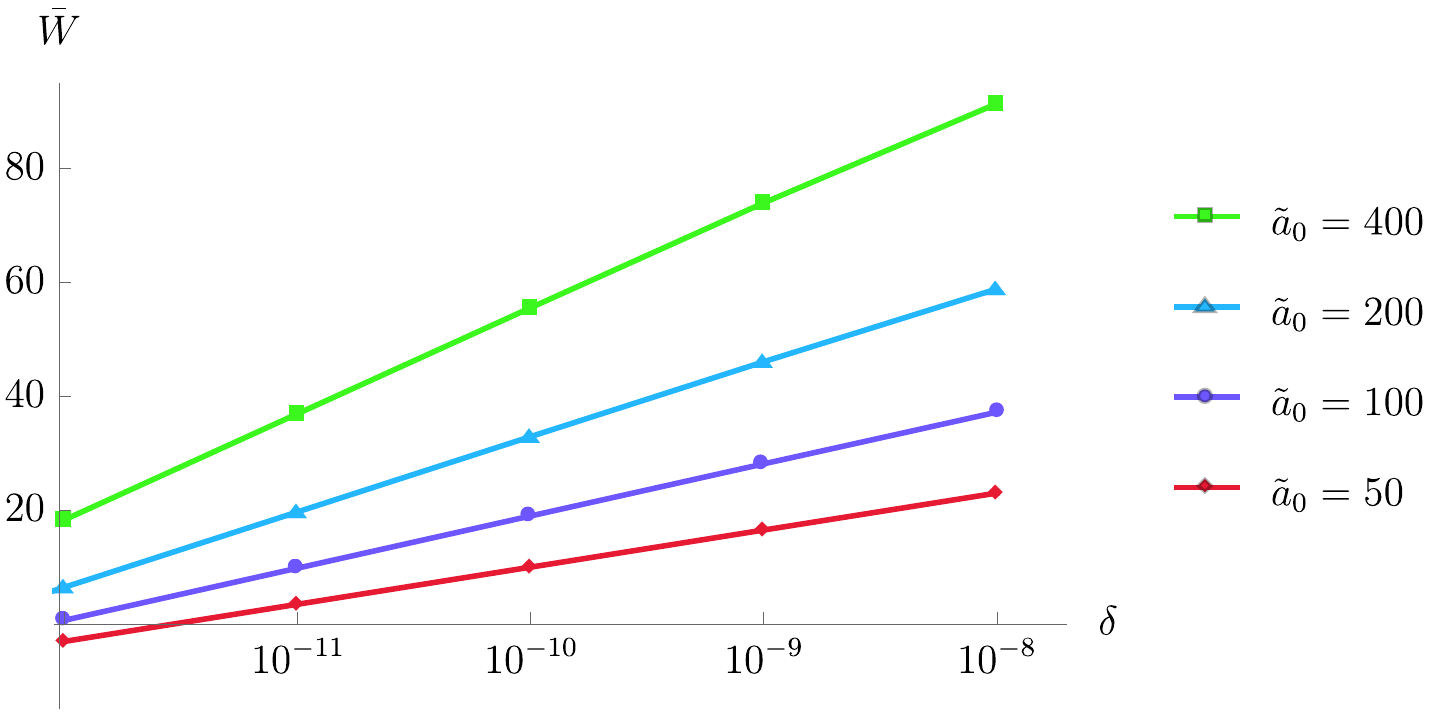}}
\caption{The suppression rate $\bar{W}$ as a function of $\delta$ and for several choices of $\tilde{a}_0$. One observes the logarithmic dependence, which excludes the possibility to treat $\delta$ as a semiclassical parameter. }
\label{Fig:DD_Kappa1}
\end{center}
\end{figure}

The net contribution of the short-distance part of the instanton is determined by a balance between the source term coming with the negative sign in eq. (\ref{WOnShell}) and the positive quartic derivative term. As figure \ref{Fig:DD_Delta} demonstrates, the difference between the two terms can be of either sign. Having confined ourselves in the region of parameters for which this difference is positive, one can try to amplify $\bar{W}$ by the means of some small constant justifying the SPA. An obvious candidate for such a constant is the parameter $\delta$ appearing in the quartic derivative term. However, from eqs. (\ref{NewAs}) and (\ref{LdLV}) it follows that
\begin{equation}\label{Ld}
\left.\bar{\L}_\delta\right\vert_{r\lesssim\bar{r}}\sim M_P\delta^{-1/6} \; .
\end{equation}
From this and eqs. (\ref{Size}) and (\ref{DD_Rho(0)}) one now sees that $\bar{W}$, in fact, does not contain any power-like dependence on $\delta$.\footnote{This fact remains true if the quartic derivative operator is replaced by an operator with a higher degree of the derivative of $\rho$, or by a linear combination thereof, see appendix \ref{AppA} for details. } This can also be inferred from figure \ref{Fig:DD_Delta}, where the dependencies of the instanton action $\bar{S}$ and of the maximum value of the radial field $\rho(0)/M_P$ on $\delta$ are shown. 

\begin{figure}[h]
\begin{center}
\center{\includegraphics[width=0.6\linewidth]{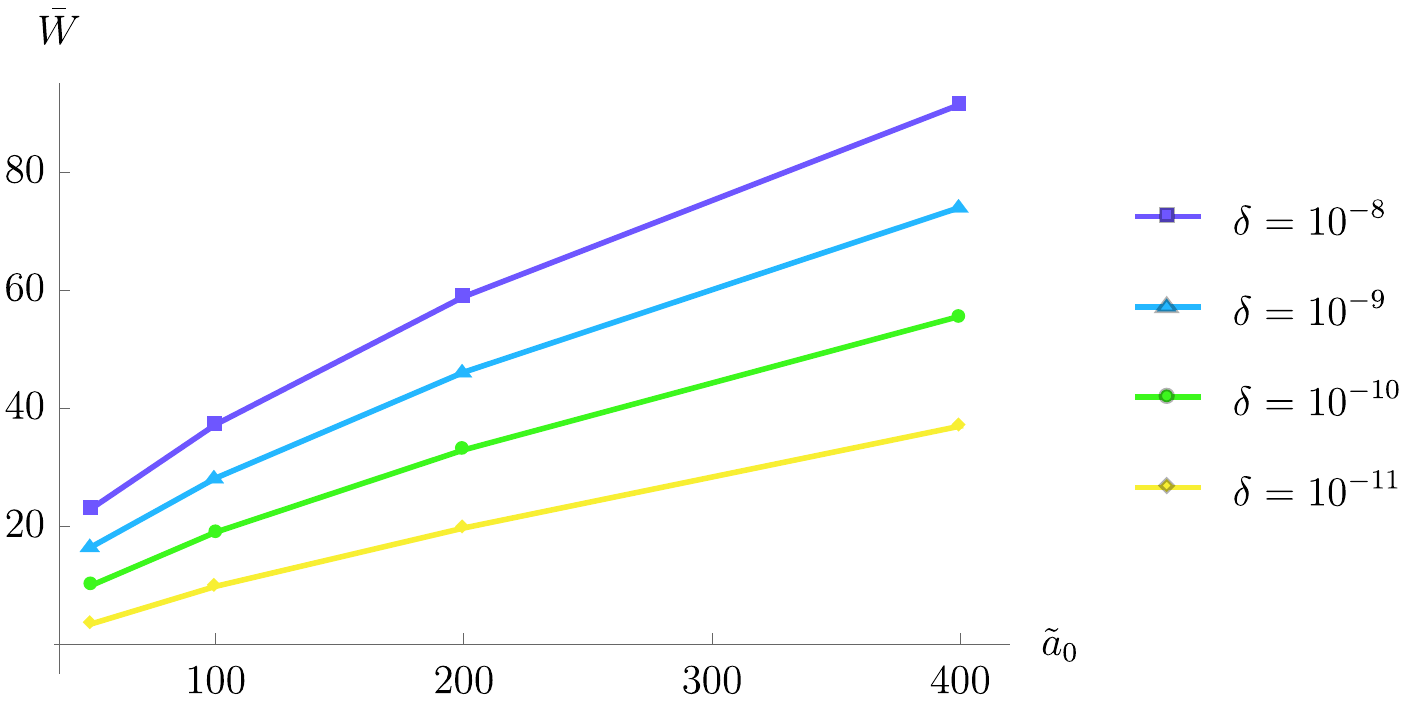}}
\caption{The suppression rate $\bar{W}$ as a function of $\tilde{a}_0$ and for several choices of $\delta$. One observes the power-like behavior, in agreement with eq. (\ref{WOnA}). The small deviations from the power law are due to a sub-dominant dependence on $\tilde{a}_0$ and an imperfect separation of the region where $\tilde{a}(\theta)$ varies from the region where the quartic derivative term dominates.  }
\label{Fig:DD_Kappa2}
\end{center}
\end{figure}

It turns out that the suitable semiclassical parameter can be provided by the asymptotics $\tilde{a}_0$ of $\tilde{a}(\theta)$. Indeed, from eqs. (\ref{Size}), (\ref{DD_Rho(0)}) and (\ref{LdLV}) one obtains that\footnote{We made use of the fact that the contribution of the singular instanton to $W$ outside the large-$\rho$ region is negligible. In what follows, this will remain true.}
\begin{equation}\label{WOnA}
\bar{W}\sim \sqrt{\tilde{a}_0} \; .
\end{equation}
As was discussed in section \ref{Ssec:NonzeroKappa}, in the models under consideration the large $\tilde{a}_0$ can be achieved by choosing the parameter $\varkappa$ to lie close to the bound in eq. (\ref{CondOnKappa}). In this case, $\tilde{a}_0^{-1}$ is the desired small parameter arising when computing the instanton value of $W$. In figures \ref{Fig:DD_Kappa1} and \ref{Fig:DD_Kappa2} we study the behavior of $\bar{W}$ as $\delta$ and $\tilde{a}_0$ vary. While the dependence on $\delta$ is seen to be logarithmic, in accordance with eq. (\ref{DD_Rho(0)}), the dependence on $\tilde{a}_0$ is power-like and matches the analytical estimation (\ref{WOnA}). Note also that eq. (\ref{WOnA}) is valid assuming that the length scale $r\sim\bar{r}$ at which the quartic derivative operator becomes dominant is smaller than the characteristic length at which the function $\tilde{a}(\theta)$ changes, and it is this fact that enabled us to replace the latter by the asymptotic value $\tilde{a}_0$ in eq. (\ref{Size}).

\section{Implications for the hierarchy problem}
\label{Sec:Pheno}

In this section, we apply the results of our analysis to the Higgs-Dilaton theory proposed in \cite{Shaposhnikov:2008xb} and studied in detail in \cite{GarciaBellido:2011de,Bezrukov:2012hx}. The part of the theory, comprising the metric, the dilaton and the Higgs fields matches the models of section \ref{Sec:DD} after the higher-dimensional operators containing the parameters $\varkappa$ and $\delta$ are introduced. As we will see, these operators do not spoil any phenomenological consequences of the theory. Within the Higgs-Dilaton model modified in this way, we demonstrate how the hierarchy between the Fermi and the Planck scales can emerge from the non-perturbative gravitational effects.

\subsection{Outline of the Higgs-Dilaton theory}
\label{Ssec:HD_Outline}

\subsubsection{Classical Lagrangian}
\label{Sssec:HD_Lagr}

The Higgs-Dilaton theory represents a moderate extension of the SM and General Relativity and possesses no dimensional parameters at the classical level. The attractiveness of the theory is due to its ability to explain certain cosmological observations as well as to provide some input into theoretical puzzles of particle physics. In particular, as we will see shortly, scale symmetry allows to reformulate the hierarchy problem (\ref{TheProblem}) in terms of dimensionless quantities. The theory naturally incorporates the Higgs inflation scenario \cite{Bezrukov:2008ut, GarciaBellido:2008ab,Bezrukov:2014ipa}, hence it predicts a successful inflationary period followed by a graceful exit to the hot Big Bang theory. Matching predictions of the theory with observational data constrains possible values of its parameters.

The Higgs-Dilaton sector of the theory becomes a particular case of the Lagrangian (\ref{DD_GenLagr_J}) upon the identification of $\varphi_1$ with the dilaton field $\chi$, and $\varphi_2$ with the dof $h$ of the Higgs field $\phi$ in a unitary gauge, $\phi^T=(0,h/\sqrt{2})$. It is written as
\begin{equation}\label{Lagr_HD_J}
\dfrac{\L_{\chi,\phi}}{\sqrt{g}}=-\dfrac{1}{2}(\xi_\chi\chi^2+2\xi_h\phi^\dag\phi)R+\dfrac{1}{2}(\d\chi)^2+\dfrac{1}{2}(\d\phi)^2+V(\chi,\phi^\dag\phi) \; ,
\end{equation}
where $(\d\phi)^2\equiv\nabla_\mu\phi\nabla^\mu\phi^*$ and the potential is given by
\begin{equation}\label{Potential_HD}
V(\chi,\phi^\dag\phi)=\l\left(\phi^\dag\phi-\dfrac{\a}{2\l}\chi^2\right)^2+\b\chi^4 \; .
\end{equation}
The full Lagrangian of the theory is obtained from eq. (\ref{Lagr_HD_J}) by supplementing the latter with the rest of the SM content. Since we are interested in the singular instanton built from $g_{\mu\nu}$, $\chi$ and $\phi$, we can ignore the presence of other dof when computing the vev of the Higgs field in the leading-order SPA. We will comment on the inclusion of other fields later in this section.

\begin{figure}[t]
\begin{center}
\center{\includegraphics[width=0.45\linewidth]{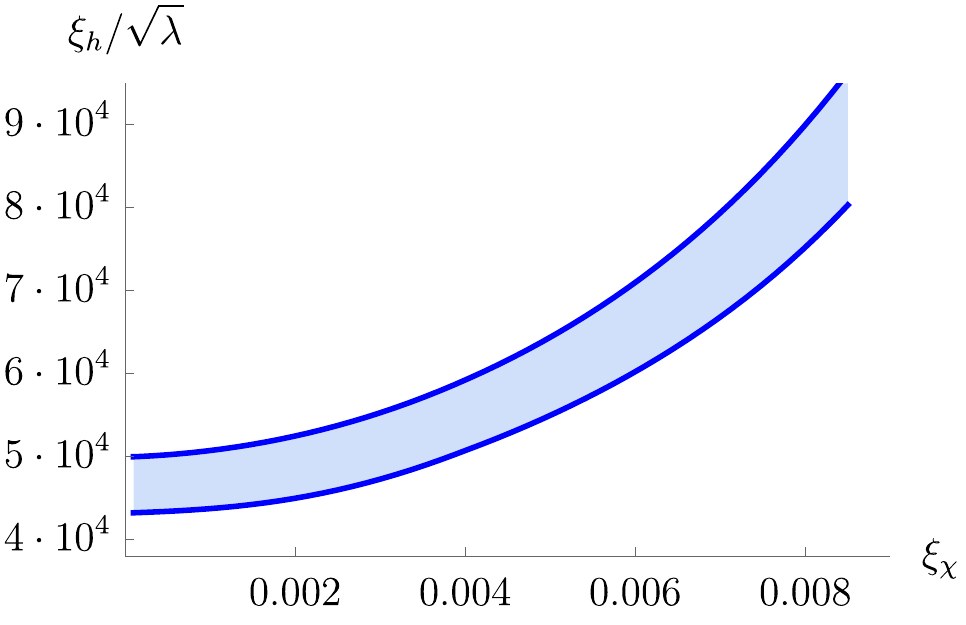}}
\caption{The parameter region for which the amplitude and the tilt of the scalar spectrum lie in the observationally allowed region, see \cite{GarciaBellido:2011de} for details. }
\label{Fig:HD_Couplings}
\end{center}
\end{figure}

As was mentioned above, the space of parameters of the theory is subject to phenomenological constraints. In particular, the values of the non-minimal couplings $\xi_\chi$ and $\xi_h$ are restricted by inflationary data. Specifically, they are bounded from measurements of the amplitude and the tilt of the primordial scalar spectrum. In figure \ref{Fig:HD_Couplings} the allowable region for $\xi_\chi$ and $\xi_h$ is shown, according to \cite{GarciaBellido:2011de}. The precise form of this region depends on details of post-inflationary processes; however, in any case
\begin{equation} \label{BoundsOnCouplings}
\xi_\chi\ll 1\ll \xi_h \; .
\end{equation}
This is different from the case studied in section \ref{Sec:DD}, where we chose the non-minimal couplings to be of the order of one. By doing so, we wanted to avoid discussing possible new interaction scales appearing in the model containing large or small couplings. It is known that in the Higgs-Dilaton model eq. (\ref{BoundsOnCouplings}) ensures the presence of additional scales, the smallest one being of the order $\sim M_P/\xi_h$. The latter, however, is still too large compared to the weak scale, hence our analysis remains in force.

The parameters $\a$, $\b$ and $\l$ in the potential (\ref{Potential_HD}) determine the low energy physics around the ground state of the theory. The latter is specified by the constant values of the dilaton and Higgs fields, $(\chi_0,h_0)^T$, where $\chi_0$ can be chosen arbitrarily and
\begin{equation}\label{GroundState_HD}
h_0^2=\dfrac{\alpha}{\lambda}\chi_0^2+\dfrac{\xi_h}{\lambda}R \; , ~~~R=\dfrac{4\beta\lambda\chi_0^2}{\lambda\xi_\chi+\alpha\xi_h} \; .
\end{equation}
The values of $\alpha$ and $\beta$ are converted into the ratios between different scales present in the SM and gravity. For example, exploiting the ratio between the Higgs and Planck masses, one obtains\footnote{In this and the following estimates we take $\lambda$ equal its low energy value, $\lambda\sim 10^{-1}$, and assume the cosmological constant to be sufficiently small.}
\begin{equation} \label{BoundOnA}
m_H^2\sim\dfrac{\alpha M_P^2}{\xi_\chi} ~~~\Rightarrow ~~~ \alpha\sim 10^{-34}\xi_\chi \; ,
\end{equation}
where the Planck mass is defined as
\begin{equation} \label{PlanckMass}
M_P^2\equiv\xi_\chi \chi_0^2+\xi_h h_0^2 \; .
\end{equation}
For the hierarchy between the Planck scale and the observed value of the cosmological constant $\Lambda$, we have
\begin{equation} \label{BoundOnB} 
\Lambda\sim\dfrac{\b M_P^4}{\xi_\chi^2} ~~~\Rightarrow ~~~ \b\sim 10^{-56}\a^2 \; .
\end{equation}
We see that the constraints (\ref{BoundOnA}) and (\ref{BoundOnB}) both involve big numbers. They are nothing but reformulations in the Higgs-Dilaton setting of the hierarchy problem (\ref{TheProblem}) and the cosmological constant problem accordingly.

The geometry of the classical ground state (\ref{GroundState_HD}) is not flat unless $\beta=0$. The non-vanishing cosmological constant required by phenomenology represents one more distinction of the Higgs-Dilaton theory from the models of section \ref{Sec:DD}. Here we appeal to the analysis of the singular instanton in the curved background performed withing the Dilaton model in appendix \ref{AppZ}. It follows that, as long as eq. (\ref{BoundOnB}) is satisfied, one can safely neglect the curvature of the background.\footnote{Note also that the Higgs-Dilaton model allows the formulation in which $\beta$ vanishes but the cosmological constant does not \cite{GarciaBellido:2011de}. }

\subsubsection{Quantum corrections}
\label{Sssec:HD_Corr}

Let us discuss the quantum corrections to the Lagrangian (\ref{Lagr_HD_J}). We choose to regularize the model in the way that makes all loop diagrams finite and all symmetries of the classical action intact. Note that the Higgs-Dilaton theory is not renormalizable \cite{Bezrukov:2012hx} (see also \cite{Shaposhnikov:2009nk}), hence an infinite number of counter-terms with the structure different from that appearing in eq. (\ref{Lagr_HD_J}) is required to be added at the quantum level. Non-renormalizability of the model does not pose a principal obstacle to its quantization, but its UV behavior cannot be uniquely fixed by the initial classical Lagrangian. The ambiguity in the choice of a set of subtraction rules is not fully removable, since it reflects our ignorance about the proper set of rules established by an unknown UV completion of the theory. Nevertheless, the underlying assumptions about the full theory, including the symmetry arguments, can constrain significantly the set of possible renormalization prescriptions. 

With the aim to preserve {\em explicitly} the scale symmetry of the theory (\ref{Lagr_HD_J}) at the perturbative quantum level, a SI renormalization procedure was developed in \cite{Shaposhnikov:2008xi} (see also \cite{Englert:1976ep} for the original suggestion and \cite{Tamarit:2013vda,Ghilencea:2016ckm,Ghilencea:2016dsl} for further developments). It is based on dimensional regularization. The use of the latter is motivated by the well-known fact that loop corrections computed within this scheme are polynomial in masses and coupling constants \cite{tHooft:1973mfk}. Hence, in the absence of heavy particle's mass thresholds, no large corrections to the Higgs mass are generated.

As an example, consider the renormalization of the Higgs self-coupling $\l$. In $d$ dimensions, one has
\begin{equation}
\l=\mu^{2\epsilon}\left(\tilde{\l}+\sum_{n=1}^\infty\dfrac{a_n}{\epsilon^n}\right) \; , ~~~ d=4-2\epsilon \; ,
\end{equation}
where by $\tilde{\l}$ we denote the dimensionless finite coupling, $\mu$ is a 't Hooft-Veltman normalization point \cite{tHooft:1972tcz} with the dimension of energy, and the series in $\epsilon$ corresponds to counter-terms. We now replace the scale $\mu$ by a field-dependent normalization point,
\begin{equation} \label{RenormScheme}
\mu^2 = F(\chi,h)\hat{\mu}^2 \; .
\end{equation}
The function $F$ reflects the particular choice of the renormalization prescription and leads to different physical results, while the dimensionless parameter $\hat{\mu}$ plays the role of the usual choice of momentum scale in the RG equations and should disappear in the final result. The scheme (\ref{RenormScheme}) is manifestly SI, as soon as $\mu$ depends only on the fields $h$, $\chi$. The change of the choice of the function $F$ can be compensated by the change of the classical Lagrangian by adding a specific set of higher-dimensional operators.  Among many possibilities, the most natural one is to identify the normalization point with the gravity scale (the first prescription) or with the SI direction along the dilaton field (the second prescription),
\begin{equation}\label{HD_Prescr}
F_I(\chi,h)= \xi_\chi\chi^2+\xi_hh^2 \; , ~~~ F_{II}(\chi,h)= \xi_\chi\chi^2 \; .
\end{equation}

Let us now discuss the quantum corrections to the Higgs mass produced in the SI scheme (\ref{RenormScheme}). It can be shown that potentially dangerous corrections from the dilaton field of the form $\l^n \chi_0^2$ cannot be generated in any order of perturbation theory \cite{Shaposhnikov:2008xi}. In particular, at one-loop level the dilaton contribution is of the form $\delta m_H^2\sim\a^2\chi_0^2$ and can be neglected in view of the constraint (\ref{BoundOnA}) and eq. (\ref{PlanckMass}). We conclude that scale symmetry makes the Higgs mass stable against radiative corrections produced by the dilaton field. Note also that in the limit $\a=0$ the dilaton decouples from the SM sector and provides no contribution to $m_H$.

The corrections to the Higgs potential from the various SM fields are well-known. They cause the spontaneous breaking of scale invariance of the tree-level Higgs potential \cite{Coleman:1973jx}. The momentum scale $\hat{\mu}$ can be chosen so that to minimize these corrections. For example, at one loop the largest contribution to the Higgs mass is provided by the top quark,
\begin{equation}
\delta m_H^2\sim m_H^2y_t^2\log\dfrac{m_t^2}{\mu^2} \; ,
\end{equation}
where $m_t^2=y_t^2h^2/2$ stands for the top quark mass. This gives,
\begin{equation} \label{Prescriptions}
\hat{\mu}^2_I=\dfrac{y_t^2}{2}\dfrac{h^2}{\xi_hh^2+\xi_\chi\chi^2} \; , ~~~ \hat{\mu}^2_{II}=\dfrac{y_t^2}{2}\dfrac{h^2}{\xi_\chi\chi^2} \; .
\end{equation}
Finally, graviton loops do not destabilize the Higgs mass as well. Indeed,  the graviton mass $m_g^2$ in the uniform $\chi_0$ and $h_0$ background (leading to the vacuum energy $\propto \lambda h_0^4$) is $m_g^2 \sim \lambda h_0^4/(\xi_\chi \chi_0^2+\xi_h h_0^2)$, and the graviton contribution to the effective potential is $\propto m_g^4$.

Let us now comment on the requirement of the absence of dof with the mass scales exceeding the EW scale. Being non-renormalizable, the Higgs-Dilaton model experiences an infinite series of counter-terms to be added to the Lagrangian (\ref{Lagr_HD_J}) in a process of renormalization. If one works at energies well below the scale at which the perturbation theory breaks down, these terms do not bring about new dof, since the particle spectrum is read from the original expression (\ref{Lagr_HD_J}).\footnote{See, e.g., chapter 16 in \cite{Hawking:1979ig}.} Then, the assumption about the absence of heavy particles amounts to the hypothesis that, as one approaches the tree-level unitarity breaking scale, the theory reorganizes itself in such a way that no undesired singularities appear in its propagators. 

\subsection{Higgs vev generation in the Higgs-Dilaton setting}
\label{Ssec:HD_GenVEV}

Let us put $\a=0$ in the potential (\ref{Potential_HD}). Then, $m_H=0$ at the classical level, according to eq. (\ref{BoundOnA}), and, in view of the discussion in section \ref{Sssec:HD_Corr}, one can be sure that the radiative corrections to the Higgs mass do not shift it towards the observed value.\footnote{We neglect the corrections to $m_H$ coming from non-zero $\b$ at the classical level.} In particular, thanks to the shift symmetry, the interaction term $\propto h^2\chi^2$ is not generated in any order of perturbation theory. Another way to see this is to notice that the RG flow of the couplings $\alpha$ and $\beta$ in the potential (\ref{Potential_HD}) is governed by
\begin{equation}\label{RG_Flow}
\mu\dfrac{d}{d\mu}\alpha=\mathcal{F}_\alpha(\alpha,\beta,...) \; , ~~~ \mu\dfrac{d}{d\mu}\beta=\mathcal{F}_\beta(\alpha,\beta,...) \; ,
\end{equation}
where $\mathcal{F}_\alpha$, $\mathcal{F}_\beta$ are functions of $\alpha$, $\beta$ and other couplings present in the theory, such that $\mathcal{F}_{\alpha,\beta}\rightarrow 0$ if both $\alpha,\beta\rightarrow 0$.\footnote{ For an equivalent discussion in terms of second-order phase transitions see \cite{Wetterich:1983bi}. } Thus, the Higgs-Dilaton theory provides a suitable framework to tackle the hierarchy problem with non-perturbative tools.

The results of section \ref{Sec:DD} are applied straightforwardly to the Higgs-Dilaton theory. In order for the mechanism to work, one must modify the theory in the limit of large magnitudes and momenta of the Higgs field. This is done by introducing the higher-dimensional operators of the form given in eqs. (\ref{DD_DefModel}). Because of their suppression by $M_P$, the vev of the Higgs field is stable against perturbative corrections coming from these operators \cite{tHooft:1974toh}. 

Following the steps performed in section \ref{Ssec:PolarVar}, we apply the Weyl rescaling to the theory (\ref{Lagr_HD_J}) to disentangle the dilaton and the Higgs fields from the Ricci scalar. We then introduce the polar field variables $\rho$ and $\theta$, and rewrite the Higgs-dilaton sector of the theory as in eq. (\ref{DD_L_NewVar}), with $a(\theta)$ replaced by $\tilde{a}(\theta)$ given in eq. (\ref{DD_aMod}). Our goal is to find numerically the singular instanton and compute its contribution to the suppression rate $\bar{W}$. 

From the results of section \ref{Ssec:NewScale} it follows that the form of the potential for the Higgs field is irrelevant for the analysis of the singular instanton. In numerical calculations we choose the potential to coincide with the RG-improved SM Higgs potential corresponding to the central values of the top quark and Higgs masses, $m_t=172.25$ GeV \cite{Castro:2017yxe}, $m_H=125.09$ GeV \cite{Aad:2015zhl}. We choose the first normalization prescription for the Higgs self-coupling $\lambda$ in eqs. (\ref{HD_Prescr}). When rewritten in terms of the polar field variables, it is given by eq. (\ref{DD_NormPoint}). We also expect the suppression rate $\bar{W}$ to be insensitive to the precise shape of the function $\tilde{a}(\theta)$ outside the vicinity of the point $\theta=\pi/2$, and, hence, to the values of the non-minimal couplings $\xi_\chi$, $\xi_h$.

\begin{figure}[t]
\begin{center}
\center{\includegraphics[width=0.75\linewidth]{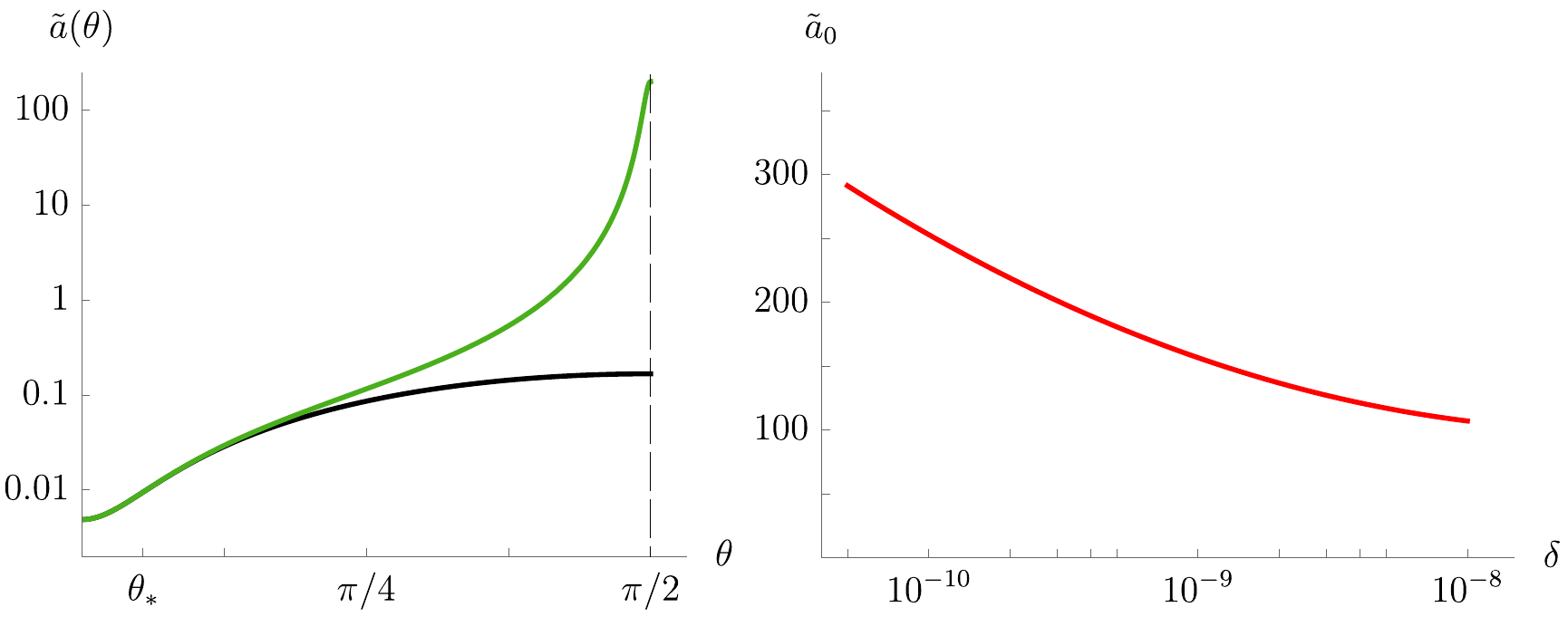}}
\caption{Left: the function $\tilde{a}(\theta)$ in the original Higgs-Dilaton theory (the lower curve) and in the modified theory with $\varkappa$ chosen so that $\tilde{a}_0=200$ (the upper curve). The angle $\theta_*$ corresponds to the scale of inflation $\sim M_P/\xi_h$. Right: the set of parameters ($\tilde{a}_0$, $\delta$), for which eq. (\ref{W0}) is satisfied. Here we choose $\xi_\chi=5\cdot 10^{-3}$, $\xi_h=5\cdot 10^3$ and $\lambda$ coinciding with the SM running Higgs self-coupling at NNLO with the central values of the top quark and Higgs masses (see figure \ref{Fig:DD_Pot}).}
\label{Fig:HD_W}
\end{center}
\end{figure}

Calculations confirm that, varying the parameters $\delta$ and $\varkappa$, one can adjust $\bar{W}$ to be equal
\begin{equation}\label{W0}
\bar{W}=\log M_P/v\approx 37 \; ,
\end{equation}
in which case the hierarchy between the Fermi and the Planck scales is reproduced in the leading-order SPA eq. (\ref{SPAScales}). This is demonstrated on the right panel of figure \ref{Fig:HD_W}.

Note that the modification of the Higgs-Dilaton theory by the higher-dimensional operators does not affect the properties which are important for phenomenology. Indeed, as the left panel of figure \ref{Fig:HD_W} demonstrates, the function $\tilde{a}(\theta)$ is indistinguishable from its counterpart in the original theory at least up to the inflationary scales. 

%This is in contrast with the results presented in figure \ref{Fig:DD_Kappa1}, where changing the value of $\varkappa$ is seen to affect $\tilde{a}(\theta)$ from relatively low scales. The difference comes from the different choice of the non-minimal couplings, expressed in eq. (\ref{BoundsOnCouplings}).

Let us comment on the dynamics of the SM dof coupled to the Higgs field. The important observation here is that the coefficient $\varkappa$ appears in the quadratic part of the Higgs field kinetic term, according to the second of eqs. (\ref{DD_DefModel}). The successful implementation of the non-perturbative mechanism requires large values of $\tilde{a}_0$, which yields $\varkappa$ to be negative. When supplementing the Higgs-gravity sector of the theory with the rest of the SM fields, one replaces the partial derivative $\d_\mu$ in the Higgs field kinetic term with the covariant one, $\mathcal{D}_\mu$. This endangers the dynamics of the gauge fields, as the latter become tachyonic as soon as they interact with the Higgs field through the SM coupling terms. This drawback can be fixed by modifying suitably the coupling of the gauge fields at high energies. For example, adding the following operator
\begin{equation}
\dfrac{(\phi\overset{\text{\scriptsize$\leftrightarrow$}}{\mathcal{D}_\mu}\phi^\dag)(\phi^\dag\overset{\text{\scriptsize$\leftrightarrow$}}{\mathcal{D^\mu}}\phi) }{2\xi_h \phi\phi^\dag+\xi_\chi\chi^2}
\end{equation}
with an appropriate coupling constant compensates the negative mass terms coming from the quadratic in derivatives operator in eq. (\ref{DD_GenLagr_J}).

\section{Discussion and conclusions}
\label{Sec:Disc}

In this paper we attempted to look at the vev of the Higgs field as arising due to some non-perturbative effect that relates the low energy phenomena with the physics at the Planck scales. We proposed that the small ratio between the Fermi and the Planck scales could be generated via the instanton configuration of a special type. We argued that in this case the Fermi scale appears as a result of an exponentially strong suppression of the Planck scale by the instanton. This effect relies strongly on a structure of the theory in the strong-gravity regime, of which explicit form we are not aware. To make possible the quantitative analysis, several conjectures about the properties of the theory in this regime were adopted. Namely, the global scale invariance was assumed to be a fundamental symmetry, broken spontaneously by the ground state associated with the Planck scale. We also assumed the absence of heavy dof associated with new physics above the weak scale. Within these conjectures, we studied several toy models comprising the gravitational and scalar fields. We constructed singular instanton configurations and investigated their contribution to the vev of the scalar field. The results of these studies were then applied to the Higgs-Dilaton theory. It was shown that the hierarchy between the Fermi and the Planck scales can indeed be generated with a particular structure of higher-dimensional operators added to the theory.

Let us summarize general features of the instanton solutions found above. The instanton action and the rate of suppression of the Planck scale in eq. (\ref{SPAScales}) are determined by the two parameters, $\tilde{a}_0$ and $\delta$, appearing in the higher-dimensional operators. From figures \ref{Fig:DD_Kappa1}, \ref{Fig:DD_Kappa2} one observes an ambiguity in the choice of these parameters leading to a given value of $W$. In particular, as figure \ref{Fig:HD_W} demonstrates, relation (\ref{W0}) can be satisfied along an entire curve in the parameter space. 
This implies that the value of $W$ reproducing the hierarchy (\ref{TheProblem}) in the leading-order SPA is not featured among other possible values. In fact, by varying $\tilde{a}_0$, one can equate $\bar{W}$ to any positive number. Thus, although the suggested mechanism allows to generate an exponentially small ratio of scales without a fine-tuning among the parameters of the theory, it does not explain a particular value of this ratio.

Speaking more generally, the mechanism is not specific to the scalar-tensor models studied in section \ref{Sec:DD}. For example, replacing the quartic derivative operator in eqs. (\ref{DD_DefModel})---(\ref{DD_DefModelPot}) by an operator with the derivatives of the scalar fields of higher degrees or by their linear combination results in the same picture. The reason is that the impact of any such operator on the short-distance behavior of the instanton is qualitatively the same. Further, due to the fact that the instanton action is saturated in the core region of the instanton, the precise shape of the function $\tilde{a}(\theta)$ regulating the strength of the radial field source is inessential, as soon as it interpolates between the given low-$\theta$ and large-$\theta$ values. Finally, including higher-dimensional operators of the types different from those considered here does not spoil the mechanism provided that they do not affect the properties of the solution near the source. As it is not so in general, we would like to stress again that, instead of performing a barely possible analysis of euclidean classical configurations arising in a general SI scalar-tensor theory of gravity, we preferred to focus on particular examples at which we demonstrate the mere possibility of the existence of the desired non-perturbative effect. 

The singular instantons found here turn out to be insensitive to the properties of the theory at low energies and low magnitudes of the Higgs field. In fact, these properties are irrelevant for the mechanism of generating the hierarchy of scales, since the latter operates essentially in the Planck region. It follows that from the perspective of a low energy theory, the vev of the Higgs field appears as a classical quantity. For example, the leading-order instanton contribution to the $n$-point correlation function of the Higgs field is given by
\begin{equation}\label{n-point}
\langle\phi(x_1)...\phi(x_n)\rangle\sim v^n \; ,
\end{equation}
provided that the points $x_1$,...,$x_n$ are farther from each other than the characteristic size of the instanton, $\vert x_i-x_j\vert\gtrsim r_*=\tilde{a}_0^{1/4}M_P^{-1}$, so that the dilute-gas approximation is applicable. Eq. (\ref{n-point}) points at the classical interpretation of the Higgs field vev, as long as the physics at the energies much below $r_*^{-1}$ is concerned. Still, there are no a priori reasons for the instanton action to be saturated exclusively in the core region of the instanton. We leave the further investigation of this question for future.

Going back to the models of section \ref{Sec:DD}, it is interesting to note that the conditions imposed on the coefficients $\delta$ and $\tilde{a}_0$ of the higher-dimensional operators point at a near Weyl-invariance of the theory in the limit $\theta\rightarrow\pi/2$. Indeed, when recast in terms of the original variables, the Lagrangian (\ref{DD_L_NewVar}) in this limit can be written as
\begin{equation}
\dfrac{\L_{\theta\rightarrow\pi/2}}{\sqrt{g}}\sim-\dfrac{1}{2}\dfrac{1}{\tilde{a}_0^{-1}-6}\varphi_2^2R+\dfrac{1}{2}(\d \varphi_2)^2+\dfrac{\delta}{1+6\xi_2}\dfrac{(\d \varphi_2)^4}{\varphi_2^4} \; .
\end{equation}
Hence, for large $\tilde{a}_0$ and small $\delta$, the theory acquires an approximate Weyl symmetry. Note again that the small coupling $\delta$, required for the mechanism to work, does not bring about new interaction scales much below $M_P$. 

It is natural to ask if singular instantons of a similar kind can be of use in resolving another great puzzle of theoretical physics --- the cosmological constant problem. Leaving the discussion of this question aside, here we just note that a straightforward attempt to implement the mechanism of section \ref{Sec:Idea} to compute the non-perturbative correction to the curvature vev $\langle R\rangle$ fails. Moreover, the scale symmetry used to make the Higgs field vev stable against radiative corrections is, in general, not suitable to protect the cosmological constant, as one can make sure using the Higgs-Dilaton theory as an example. 

As was discussed in section \ref{Sec:intro}, the global scale symmetry is a useful guiding principle in a search for theories on which possible non-perturbative quantum gravity effects can be tested. From this point of view, incorporating gravity into a theory in a SI way is advantageous. Nevertheless, the mechanism of generating the hierarchy (\ref{TheProblem}) can be successfully implemented in cases when the global scale invariance is not respected by the gravitational sector of the theory at low energies. This situation was studied in \cite{Shaposhnikov:2018xkv} using models of one scalar field coupled to gravity in a non-minimal way. In those models, a classically zero vev of the scalar field is protected against perturbative corrections by a global conformal symmetry of the scalar sector. The vev can then receive non-perturbative contribution via the singular instanton analogous to that of the Dilaton model. The analogy is due to the fact that at high energies the model enters the SI regime in which it reduces to the Dilaton model improved by suitable Planck-suppressed higher-dimensional operators.

In the language of the Higgs-Dilaton theory, our motivation in searching for a non-perturbative mechanism of generating the Higgs vev was an unnatural smallness of the coefficient $\a$ in the potential (\ref{Potential_HD}). One more parameter of the theory which is required to be small in order to match phenomenological data is the non-minimal dilaton coupling $\xi_\chi$. In the limit $\xi_\chi=0$, the Lagrangian (\ref{Lagr_HD_J}) without the potential term acquires an additional invariance under the constant shifts of the dilaton field. It is tempting to suggest that this shift symmetry is exact at the classical level, and that the interaction $\propto\chi^2R$ is induced by some non-perturbative effect similar to the one studied, e.g., in \cite{Adler:1980md,Hasslacher:1980mw,Adler:1982ri}. We leave the investigation of this appealing possibility to the future work.

\acknowledgments

The authors thank Fedor Bezrukov and Javier Rubio for helpful discussions. This work was supported by the ERC-AdG-2015 grant 694896. The work of M.S. and A.S. was supported partially by the Swiss National Science Foundation.

\appendix
\section{Singular instanton in curved space}
\label{AppZ}

Let us switch on the quartic coupling $\lambda$ in the Lagrangian (\ref{D_L_J}) of the Dilaton model. Then, the second of eqs. (\ref{D_EOM}) becomes
\begin{equation}\label{D_SingInst_1}
\dfrac{1}{f^2}=1+\dfrac{a}{6M_P^4r^4}\pm b^2r^2 \; , ~~~ b^2=\dfrac{\vert\lambda\vert M_P^2}{12\xi^2} \; ,
\end{equation}
where the plus (minus) sign in the second expression holds for negative (positive) $\lambda$. The classical ground state (\ref{D_CGS_J}) of the Dilaton model is given by
\begin{equation}\label{D_CGS_E_1}
\bar{\varphi}=0 \; , ~~~ f^2=\dfrac{1}{1\pm b^2r^2} \; ,
\end{equation}
Repeating the steps leading to eq. (\ref{D_SingInst}), we obtain the expression for the singular instanton in the space of constant curvature,
\begin{equation}\label{D_SingInst_2}
\bar{\varphi}(r)=-\int_{r_b}^r \dfrac{f(r')}{r'^3M_P} dr' \; , ~~~ \dfrac{1}{f(r)^2}=1+\dfrac{a}{6M_P^4r^4}\pm b^2r^2 \; ,
\end{equation}
where $r_b$ is sent to infinity for $\lambda<0$ or is equal to a positive root of the inverse of the metric function $f^{-2}$ for $\lambda>0$. Eq. (\ref{D_SingInst_2}) contains two scales. The first of them is defined by the combination $a^{1/4}M_P^{-1}$ and determines the size of the instanton, as explained in section \ref{Ssec:Class}. The second is the cosmological scale $b$ determined by the classical ground state. We require the vacuum energy of the ground state to be well below $M_P^4$,
\begin{equation}\label{D_CondOnLambda}
bM_P^{-1}\ll 1 \; .
\end{equation}
From this and the fact that $a$ is confined in the region
\begin{equation}\label{D_CondOnA}
0<a<1/6 
\end{equation}
the separation of the instanton and cosmological sizes follows. Eq. (\ref{D_CondOnLambda}) imposes an upper bound on the absolute value of $\lambda$, which can always be satisfied provided that $\xi\neq 0$.

It is worth to note that when the vacuum geometry is the de Sitter one, $\lambda>0$, the instanton is not regular at the boundary point $r=r_b$. However, computation of the metric invariants yields, in notations of \cite{Greenberg},\footnote{Among the fourteen metric invariants, ten are expressed using the Weyl tensor which is zero in our case \cite{Greenberg}. }
\begin{align}
\tilde{R}= 12b^2(1+\mathcal{O}(ab^4M_P^{-4})) \; , ~~~ & \tilde{E}=b^4\cdot\mathcal{O}( a^2b^{8}M_P^{-8}) \; , \\
\tilde{F}=b^8\cdot\mathcal{O}( a^4b^{16}M_P^{-16} )\; , ~~~ & \tilde{G}=b^{12}\cdot\mathcal{O}( a^6b^{24}M_P^{-24}) \; . \nonumber
\end{align}
To the leading order in $a^{1/4}bM_P^{-1}$, they coincide with those of the euclidean de Sitter space. Hence, one can expect that the singularity of the metric at $r=r_b$ does not contribute to the instanton action.

As the last step, we evaluate the euclidean action and the boundary term of the instanton in curved background. With the ansatz (\ref{Ansatz}) applied, the exterior curvature of a surface defined by the equation $r=r_s$ is seen to be 
\begin{equation}
\tilde{K}=\dfrac{3}{f(r_s)r_s} \; .
\end{equation}
For $\lambda$ positive, the boundary term is absent both for the vacuum solution and the singular instanton. In the case $\lambda < 0$, the boundary is determined by sending $r_s$ to infinity and we have (cf. eq. (\ref{Difference}))
\begin{equation}
\bar{I}_{GH}-I_{GH,0}\sim a^{-1}b^{-1}M_P^{-2}r_s^{-3}\rightarrow 0 \; , ~~~ r_s\rightarrow\infty \; , ~~~ \lambda<0 \; .
\end{equation}
To find the euclidean action, we make use of the Einstein equations. The difference in the actions between the instanton and the vacuum for $\lambda\neq 0$ is evaluated as
\begin{equation}
\bar{S}-S_0\sim a b^2 M_P^{-2} \ll 1
\end{equation}
given eqs. (\ref{D_CondOnLambda}) and (\ref{D_CondOnA}).

We conclude that the nontrivial background geometry does not lead to a significant contribution to the net instanton action, neither to the net boundary term. Hence, in proceeding with the classical analysis in more complicated theories, one can focus solely on the core region of the instanton. Moreover, as was mentioned in section \ref{Ssec:Class}, in order to make the instanton action large, the structure of the theory in this region must be different from that of the Dilaton model. 

\section{Derivative operators of higher degrees  }
\label{AppA}

Here we discuss the generalization of the models of section \ref{Sec:DD}, which amounts to replacing the quartic derivative term for the radial field by a more general operator of the form
\begin{equation}\label{A_O}
\tilde{\mathcal{O}}=\delta M_P^4p(z) \; , ~~~ p(z)=\sum_{n=1}^N\a_nz^n \; , ~~~ z=\dfrac{(\tilde{\d}\rho)^4}{M_P^8} \; .
\end{equation}
The original operator is reproduced when $p(z)=z$, $\a_1=1$. The coefficients $\a_n$ are chosen to be less or of the order of one, the overall coupling $\delta$ is adjusted to provide the separation of the region where the angular field varies from the region where the operator $\tilde{\mathcal{O}}$ dominates the dynamics of the instanton. Each of the terms in $p(z)$ can be easily traced back to the original field variables, invoking non-zero coefficients $\tilde{\gamma}_{i_1,...i_{2k}}$ up to $k=N/2$ in the Lagrangian (\ref{DD_GenLagr_J}).

Making use of the Einstein equations, one finds the instanton action 
\begin{equation}
\bar{S}=\int d^4x\sqrt{\tilde{g}}\delta M_P^4 (2zp'(z)-p(z)) \; ,
\end{equation}
where the potential term is neglected. We would like to study how this action depends on the coupling $\delta$ for different choices of the function $p(z)$. Applying the Ansatz (\ref{Ansatz}), we arrive at equations of motion in the high energy regime,
\begin{align}\label{A_EOM}
& 4r^3\delta M_P^2 z^{\frac{3}{4}}p'(z)=-\dfrac{1}{M_P} \; , \\ 
& \dfrac{M_P^2}{2}\dfrac{3-3f^2}{r^2f^2}=\delta M_P^4(2zp'(z)-p(z)/2) \; .
\end{align}
Let us take
\begin{equation}
p(z)=z^k \; , ~~~ k\geqslant 1 \; .
\end{equation}
From eqs. (\ref{A_EOM}) the high energy asymptotics of the radial and metric fields are deduced,
\begin{align}\label{A_As}
& \rho'\sim -M_P^2\delta^{\frac{1}{2-8k}}(M_Pr)^{\frac{2k-2}{4k-1}} \; ,\\
& f\sim\delta^{\frac{1}{8k-2}}(M_Pr)^{\frac{2k+1}{4k-1}} \; ,\nonumber
\end{align}
where we keep track of the dependence on $\delta$ and $\tilde{a}_0$. These asymptotics prevail at the distances $r\lesssim \bar{r}$, where
\begin{equation}\label{A_r}
\bar{r}\sim\delta^{\frac{1}{6(2k-1)}}M_P^{-1}\tilde{a}_0^{\frac{4k-1}{6(2k-1)}} \; .
\end{equation} 
Setting $k=1$, one reproduces eqs. (\ref{NewAs}), (\ref{Size}). The instanton action becomes,
\begin{equation}
\bar{S}\sim\int_0^{\infty}drr^3f\delta M_P^4\left(\dfrac{\rho'^4}{f^4M_P^8}\right)^k \; .
\end{equation}
We now use eqs. (\ref{A_As}) and (\ref{A_r}) to compute the high energy part of the action. Remarkably, it shows no power-like dependence on $\delta$:
\begin{equation}
\bar{S}\sim\tilde{a}_0^{\frac{1}{2}} \; .
\end{equation}
The same is true for the value of the radial field at the center of the instanton,
\begin{equation}
\rho(0)/M_P\sim\tilde{a}_0^{\frac{1}{2}}(\log\delta+\mathcal{O}(1)) \; .
\end{equation}

It is clear that using the more general form of the function $p(z)$, given in eq. (\ref{A_O}), reveals the same behavior of $\bar{S}$ and $\rho(0)/M_P$. We conclude that the reasoning of section \ref{Ssec:NonzeroKappa} applies universally regardless the particular derivative operator chosen to regularize the instanton. 

From eqs. (\ref{A_As}) it also follows that the high energy asymptotics of the fields are confined as
\begin{equation}
\vert\rho'\vert\gtrsim r^\frac{1}{2} \; , ~~~ r^\frac{1}{2}\gtrsim f\gtrsim r \; .
\end{equation}
Hence, the non-analyticity invoked by the source of the radial field cannot be completely removed by the operators of the form (\ref{A_O}). 

\section{More on short-distance behavior of the instanton}
\label{AppB}

\begin{figure}[b]
\begin{center}
\center{\includegraphics[width=0.45\linewidth]{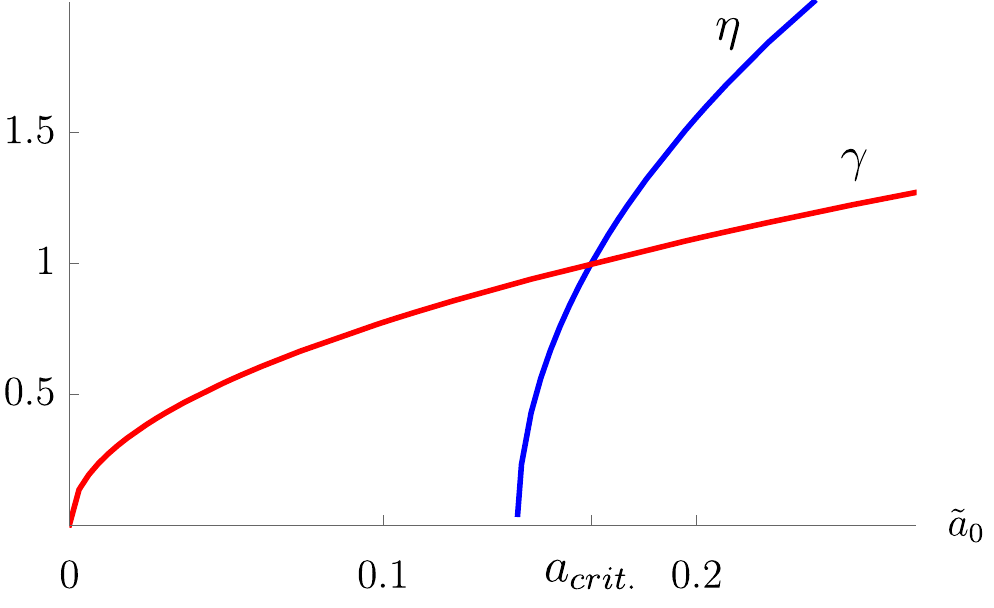}}
\caption{The exponents of the short-distance asymptotics of the fields $\rho$ and $\theta$ with no higher-dimensional derivative terms included. Here we take $\xi_1=1$, $\xi_2=1.1$. The critical value of the source of the radial field, $a_{crit.}$, is indicated according to eq. (\ref{B_aCrit}). }
\label{Fig:B}
\end{center}
\end{figure}

Following the discussion in section \ref{Ssec:NonzeroKappa}, here we study the exponents $\gamma$, $\eta$ in the asymptotics of the scalar fields at $r\rightarrow 0$ and for different values of $\varkappa$. Recall that 
\begin{equation}
\tilde{a}_0\equiv \tilde{a}(\pi/2) \; ,
\end{equation}
where $\tilde{a}(\theta)$ is a function defined in eq. (\ref{DD_aMod}). From equations of motion for the radial and angular fields it follows that
\begin{equation}
\rho\sim-M_P\gamma\log(M_Pr) \; , ~~~ \dfrac{\pi}{2}-\theta\sim r^\eta
\end{equation}
with
\begin{equation}
\gamma=\sqrt{6\tilde{a}_0} \; ,  ~~~ \eta=\sqrt{\dfrac{\tilde{a}_0(1+6\xi_2)(2\xi_2(1+3\xi_1)-\xi_1)-\xi_2^2(1+6\xi_1)}{\xi_1(\xi_1+1/6)}} \; .
\end{equation}
This reduces to eqs. (\ref{Gamma}) and (\ref{Beta}) for $\tilde{a}_0=a_0\equiv (6+1/\xi_2)^{-1}$.

Figure \ref{Fig:B} demonstrates the relative values of $\gamma$ and $\eta$ for different possible values of the coefficient $\tilde{a}_0=(\varkappa+a_0^{-1})^{-1}$. We observe two featured values of $\tilde{a}_0$. The first one represents the minimal possible strength of the source for which the singular instanton of the type studied here exists. It is given by
\begin{equation}\label{B_aMin}
a_{min.}=a_0\dfrac{\xi_2(1+6\xi_1)}{\xi_2(2+6\xi_1)-\xi_1} \; .
\end{equation}
If $\varkappa=0$, the requirement $\tilde{a}_0>a_{min.}$ gives $\xi_2>\xi_1$, in agreement with eq. (\ref{Beta}). The second featured value of $\tilde{a}_0$ is the one at which $\eta=\gamma$. It is given by
\begin{equation}\label{B_aCrit}
a_{crit.}=a_0\dfrac{\xi_2(1+6\xi_1)}{\xi_2(1+6\xi_1)-\xi_1}
\end{equation}
and is always larger than $a_0$. For $\tilde{a}_0>a_{crit.}$ we have, according to eq. (\ref{AsOldVar}),
\begin{equation}
\varphi_1\rightarrow 0 \; , ~~~ r\rightarrow 0 \; . 
\end{equation}
Thus, the large sources make the dilaton field associated with $\varphi_1$ convergent at the center of the instanton. Note, however, that the behavior of the dilaton is still non-analytic in $r$, which is justified by the presence of the source. Furthermore, the Higgs field associated with $\varphi_2$ diverges the stronger, the larger the value of $\tilde{a}_0$, hence the regularization provided, for example, by the higher-dimensional derivative operator is still needed.

\bibliographystyle{JHEP}

\bibliography{HD-instanton-Ref}

\end{document}